%% file: main.tex
\RequirePackage{fix-cm}
\documentclass[twocolumn,epjc3]{svjour3}  
\smartqed  
\RequirePackage{graphicx}

\RequirePackage{mathptmx}      
\usepackage{todonotes}
\usepackage{siunitx}
\usepackage{amsmath}
\usepackage{amssymb}
\usepackage[switch,mathlines]{lineno}
\usepackage{xspace}
\usepackage[T5,T1]{fontenc}
\usepackage[compact]{titlesec}
\RequirePackage[colorlinks,citecolor=blue,urlcolor=blue,linkcolor=blue]{hyperref}
\journalname{Eur. Phys. J. C}
\let\oldequation\equation
\let\oldendequation\endequation
\renewenvironment{equation}
  {\linenomathNonumbers\oldequation}
  {\oldendequation\endlinenomath}

\begin{document}


\title{Observation of Seasonal Variations of the Flux of High-Energy Atmospheric Neutrinos with IceCube
}

\titlerunning{Seasonal Variations of High-Energy Atmospheric Neutrinos}        
\include{authorlist}




\date{Received: date / Accepted: date}

\maketitle

\begin{abstract}
Atmospheric muon neutrinos are produced by meson decays in cosmic-ray-induced air showers.
The flux depends on meteorological quantities such as the air temperature, which affects the density of air. Competition between decay and re-interaction of those mesons in the first particle production generations gives rise to a higher neutrino flux when the air density in the stratosphere is lower, corresponding to a higher temperature. 
A measurement of a temperature dependence of the atmospheric $\nu_{\mu}$ flux provides a novel method for constraining hadro\-nic interaction models of air showers. It is particularly sensitive to the production of kaons. Studying this temperature dependence for the first time requires a large sample of high-energy neutrinos as well as a detailed understanding of atmospheric properties.  \newline
We report the significant ($> 10 \; \sigma$) observation of a correlation between the rate of more than 260,000 neutrinos, detected by IceCube between 2012 and 2018, and atmospheric temperatures of the stratosphere, measured by the Atmospheric Infrared Sounder (AIRS) instrument aboard NASA's AQUA satellite. For the observed 10$\%$ seasonal change of effective atmospheric temperature we measure a 3.5(3)$\%$ change in the muon neutrino flux. This observed correlation deviates by about 2-3 standard deviations from the expected correlation of 4.3$\%$ as obtained from theoretical predictions under the assumption of various hadronic interaction models. 
\end{abstract}

\section{Introduction}
\label{sec:Intro}

\begin{figure}
    \centering
    \includegraphics[width = 0.98\columnwidth]{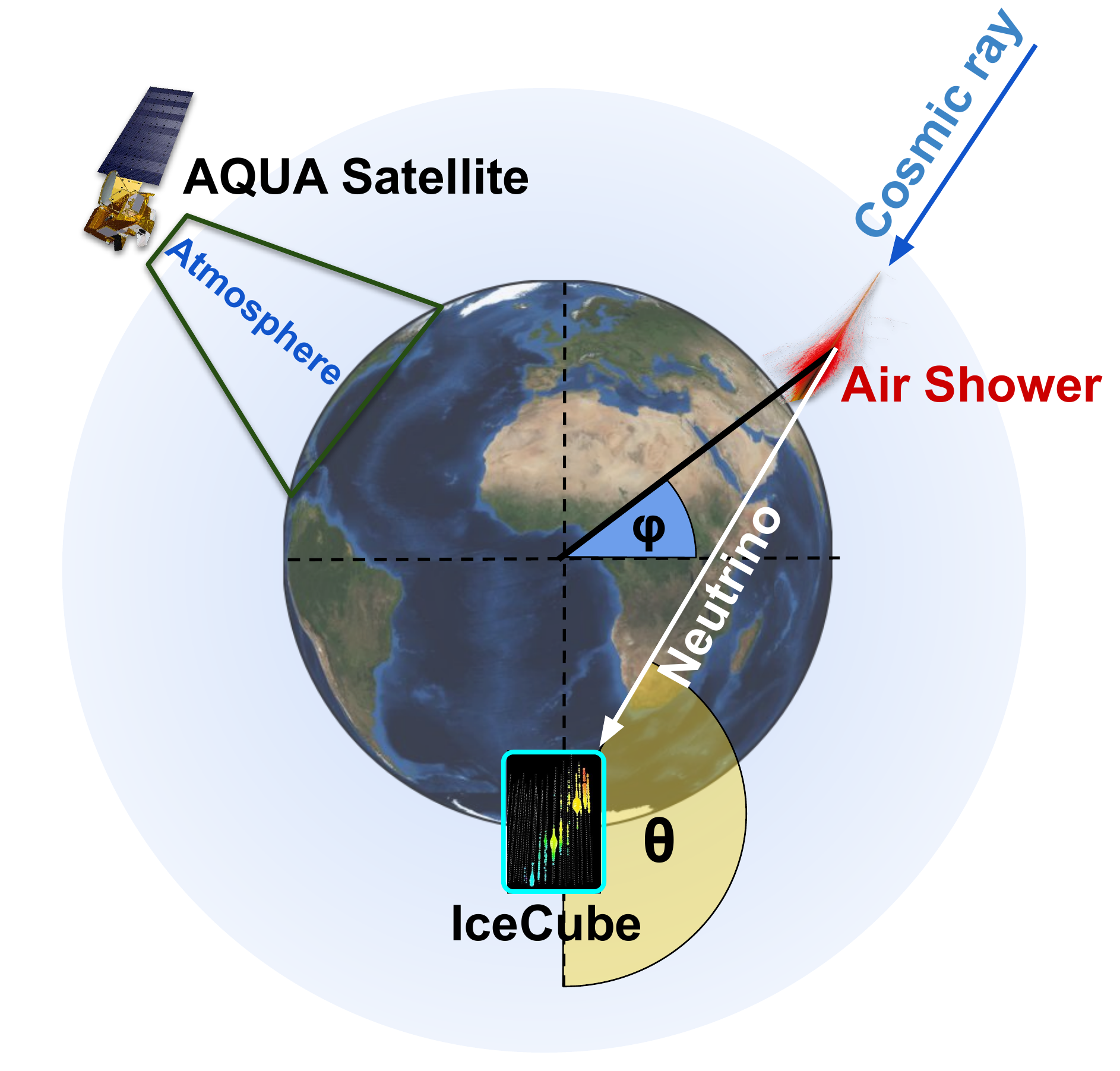}
    \includegraphics[width = 0.98\columnwidth]{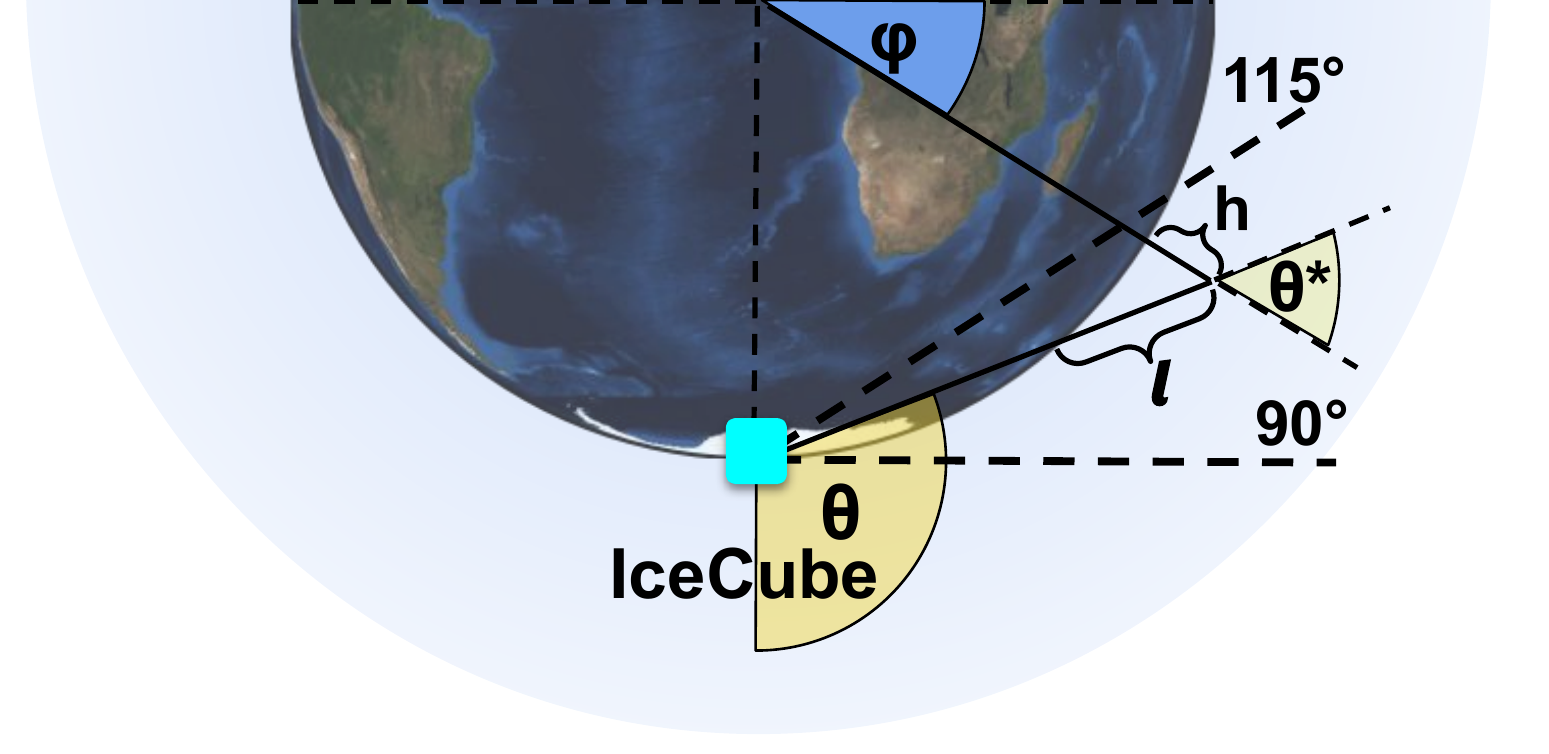}
    \caption{
    Sketches of the experimental setup. The top sketch shows in the right half the production, propagation and measurement of an atmospheric $\nu_{\mu}$ with IceCube. It also shows in the top left the measurement of the temperature by the AIRS instrument. The observation zone (IceCube zenith from \SIrange{90}{115}{\degree}) can be seen in the bottom sketch. This sketch also contains the definition of geometrical quantities such as $h$, $l$ and $\theta^*$. These quantities are explained further in
    \ref{sup:integrate}.
    The sketch is not to scale.
    }
    \label{fig:NeutrinoPath}
\end{figure}

Atmospheric muon neutrinos are produced in weak decays of mesons that are produced in cosmic-ray-induced air showers in Earth's atmosphere~\cite{Gaisser:2002jj}.
These mesons either decay and produce neutrinos and muons or interact with air. The probability of interaction decreases proportional to the local density of air, which is inversely proportional to the atmospheric temperature~\cite{GaisserBook}.
When the probability for mesons to interact decreases, the probability of decay and the subsequent production of atmospheric neutrinos and muons increases. Therefore, the relative change in flux of atmospheric neutrinos and muons is expected to correlate with the relative change in the temperature of Earth's atmosphere. 

Locally, the atmospheric temperature changes continuously resulting in daily variations. Additionally, these small variations are modulated by
a larger yearly variation that is related to the seasonal change of the global atmosphere. 
For atmospheric muons, which arise from weak decays of mesons as well, similar variations are a well-established observation, see e.g.\ \cite{Barrett:1952woo,Tilav:2019xmf,bouchta1999seasonal,Tilav:2010hj,MINOS:2009njg,Adamson:2014xga,DoubleChooz:2016sdt,OPERA:2018jif,NOvA:2019rnr,MACRO:1997teb,Sagisaka:1986bq,LVD:2019zlh,LVD:2022yvo}.
For atmospheric $\nu_{\mu}$ a similar measurement is challenging because of the large number of events required for a significant observation. Additionally, the flux of atmospheric $\nu_{\mu}$ is observed from all zenith directions due to these neutrinos penetrating Earth. This requires the measurement of temperatures globally unlike the atmospheric muon flux which requires the observation of temperatures at the experiment's location.

The production of high-energy atmospheric neutrinos and muons on average occurs at high altitudes
 of ~\SIrange{20}{40}{km} (corresponding to a range from \SIrange{3}{60}{\hecto \pascal}) where the atmospheric density is sufficiently large for the production of the parent mesons, but still sufficiently small for these mesons to decay.
Therefore, the measurement of the correlation of atmospheric $\nu_{\mu}$ with the stratospheric temperature probes the meson production
in the early development of cosmic-ray air showers at high altitude.
Unlike the flux of atmospheric muons, which is generally dominated by decays of charged pions, the flux of high-energy atmospheric $\nu_{\mu}$ has a substantial contribution from decays of charged kaons~\cite{Gaisser:2002jj,Desiati:2010wt}. This is caused by the different energy fraction dependencies in meson decays for neutrinos ($1 - m_{\mu}^2/m_{K/\pi}^2$) and muons ($m_{\mu}^2/m_{K/\pi}^2$). Due to the higher mass, energy fractions of muons and neutrinos in kaon decays are much closer compared to pion decays.
This contribution amounts to roughly \SI{30}{\percent} at lower neutrino energies ($\lesssim \SI{10}{GeV}$) but increases with neutrino energy reaching about \SI{80}{\percent} at high neutrino energy ($\gtrsim \SI{1}{TeV} $) (see \cite{MCEqpaper}).
 The production of kaons at high energy is still a major uncertainty in the modeling of air showers and atmospheric $\nu_{\mu}$ fluxes~\cite{BarrPaper,Fedynitch:2017trv} and the analysis of seasonal variations provides a new opportunity for constraining such uncertainties.
 
The IceCube Neutrino Observatory~\cite{IceCubeDetectorPaper} is measuring high-energy neutrinos of astrophysical origin. While IceCube was designed to measure astrophysical neutrinos, it also detects a unique data sample of high-energy atmos\-pheric $\nu_{\mu}$ with unprecedented statistics of several hundred thousand events in ten years at energies above \SI{100}{GeV} \cite{IceCube:2021uhz}.
Preliminary studies of seasonal variations of the $\nu_{\mu}$ flux in IceCube have been reported earlier \cite{Gaisser:2013lrk,Heix:2019jib}.
  
A sketch of the analysis concept is shown in Fig. \ref{fig:NeutrinoPath}.
For each detected neutrino in IceCube, the reconstructed direction is used to determine the atmospheric temperature profile at the production site.
The Atmospheric Infrared Sounder (AIRS) on NASA's Aqua satellite~\cite{AIRSReference} provides a daily coverage of the global distribution of atmospheric temperatures. 
This atmospheric temperature dataset allows for a description of the atmospheric conditions for the respective time and direction of each neutrino event.
The data is analyzed two ways: using an un-binned likelihood approach as well as a $\chi^2$ fit with bins of daily temperature and neutrino rates. 
The results are investigated for various systematic effects and compared to predictions using the numerical cascade equation solver MCEq~\cite{MCEqpaper} with cosmic-ray primary particle fluxes, and different state-of-the-art hadronic interaction models.

\section{Data sets}
\label{sec:datasets}

\subsection{The IceCube Neutrino Observatory}
\label{sec:IceCube}

\begin{figure*}
    \centering
    \includegraphics[width =0.85\textwidth]{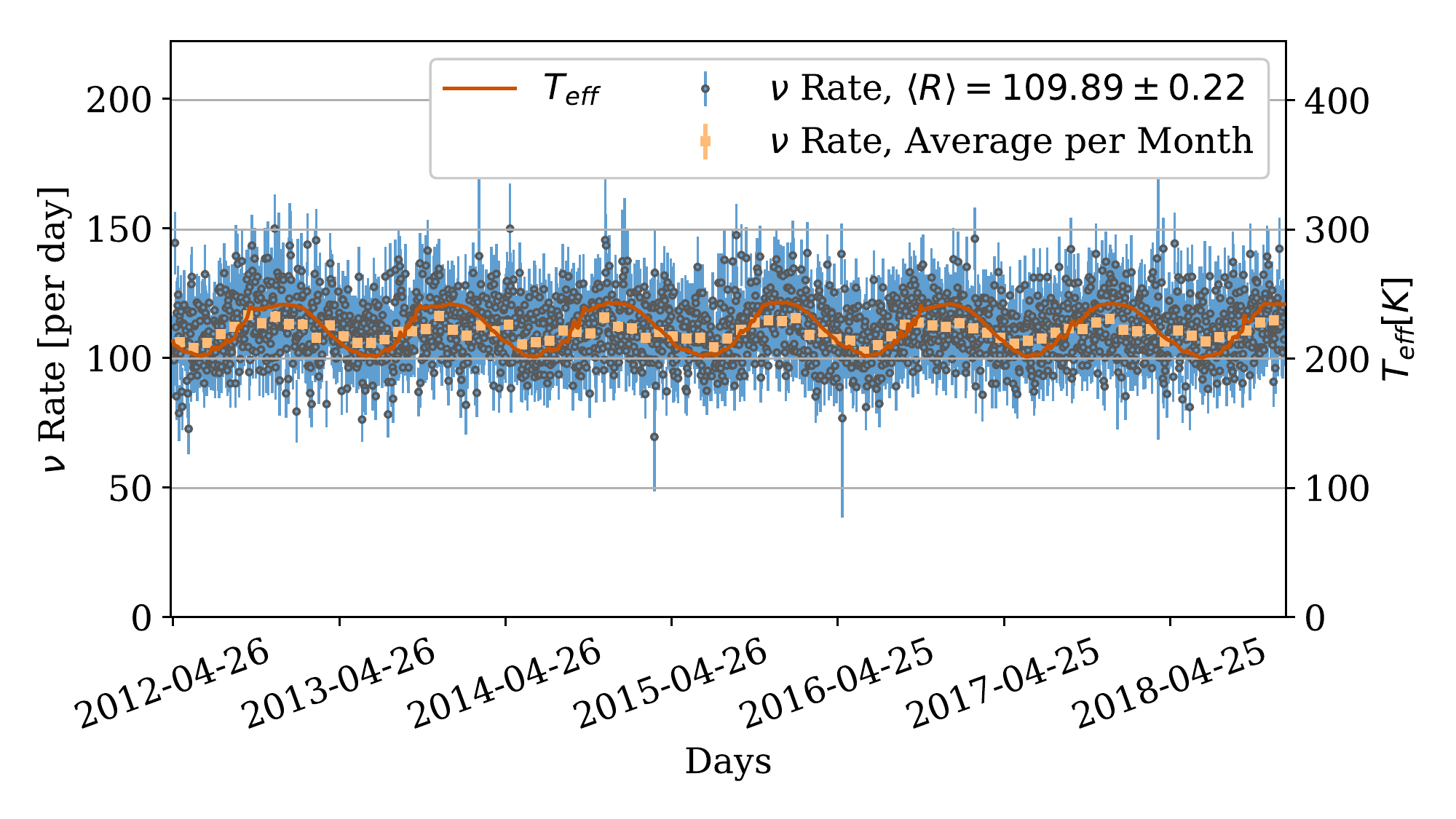}
    \caption{Plot showing the data used in the seasonal variations analysis. The neutrino data is shown in daily bins (blue points/error bars), monthly bins (orange points/error bars). The red line depicts the effective temperature as defined in Eq.\ref{eq:TeffDef} based on the data measured by AIRS. The temperature values are on the right hand y-axis of the plot.}
    \label{fig:daily_month_rate_temp}
\end{figure*}

The IceCube Neutrino Observatory~\cite{IceCubeDetectorPaper} is a cubic-kilometer Cheren\-kov detector located at the geographic South Pole.
It detects neutrino interactions by measuring Cheren\-kov light from secondary charged particles using 5160 digital optical modules (DOMs) buried in the antarctic ice at depths of \SIrange{1450}{ 2450}{m} below the ice's surface. Each DOM consists of a hemispherical photomultiplier tube which is embedded in a glass pressure housing together with electronics for digitization, control and communication to the surface. These DOMs 
are placed along 86 vertical strings arranged in a hexagonal structure.
This analysis uses a sample of through-going muon tracks induced by up-going and horizontal muon-neutrinos which has been recorded between April 2012 and December 2018 , corresponding to 2443 days  
of data-taking and an uptime of about  97\%. This uptime includes further quality requirements compared to the IceCube duty-cycle.

The data selection is identical to the analysis in~\cite{IceCube:2021uhz} which investigates the diffuse astrophysical neutrino spectrum.
The sample uses the Earth as a natural shield (zenith range from \SIrange{85}{180}{\degree}) to suppress atmospheric muons and additional cuts are applied on the goodness of the events reconstructions and energy deposited in the detector.
The selection reaches a purity of neutrino-induced muons of ~99.85\%, estimated using simulations. 
In addition to the criteria described in \cite{IceCube:2021uhz}, we narrow the field of view to neutrinos  reconstructed within zenith directions from \SIrange{90}{115}{\degree}.
Neutrinos in this range originate dominantly from air showers in  Earth's Southern hemisphere with geographic latitudes between \SIrange{-90}{-40}{\degree}, as shown in Fig. \ref{fig:NeutrinoPath}.
With this choice, we exclude neutrinos from geographic regions of opposite seasons
(Northern hemisphere) and regions with small seasonal temperature variations close to the Earth's equator.
The resulting sample consists of
a total of \num{262 846} events. This corresponds to an average of \num{110} detected neutrinos per day. The measured  daily rates of neutrinos are shown in Fig. \ref{fig:daily_month_rate_temp}.
Based on the fit of the measured energy spectrum in \cite{IceCube:2021uhz}, we expect a median neutrino energy of \SI{900}{GeV}, and \SI{90}{\percent} of neutrinos to have energies between \SI{200}{GeV} and \SI{7 700}{GeV}.

\subsection{The Atmospheric Infrared Sounder AIRS}
\label{sec:AIRS}
The measured neutrino rate is correlated with the atmospheric temperature at the geographic location of the parent air shower. The temperature data required by this analysis is obtained by the Atmospheric Infrared Sounder (AIRS) \cite{AIRSReference}. The satellite
orbits Earth on a Sun-synchronous orbit,
crossing the equator at about 13:30h and 1:30h local time.
AIRS observes a swath of \SI{1650}{km} and gives an angular resolution of \SI{1}{\degree}$\times$\SI{1}{\degree} in latitude and longitude. 
This results in observational gaps around the equator (see Fig. \ref{fig:TempDataTaking}) but a good overlapping coverage of the atmosphere in the considered geographic range
with at least two observations per day for every location.
The AIRS data product is publicly available \cite{AIRSReference}.
For each measurement the data include temperatures on 24 isobar levels ranging from \SIrange{0.1}{1000}{hPa}, and additionally include the altitude of each of these pressure levels. The combined accuracy is given as \SI{1}{K} per \SI{1}{km} of vertical depth \cite{AIRSReference}. 
\begin{figure}
    \centering
    \includegraphics[width = 0.45\textwidth]{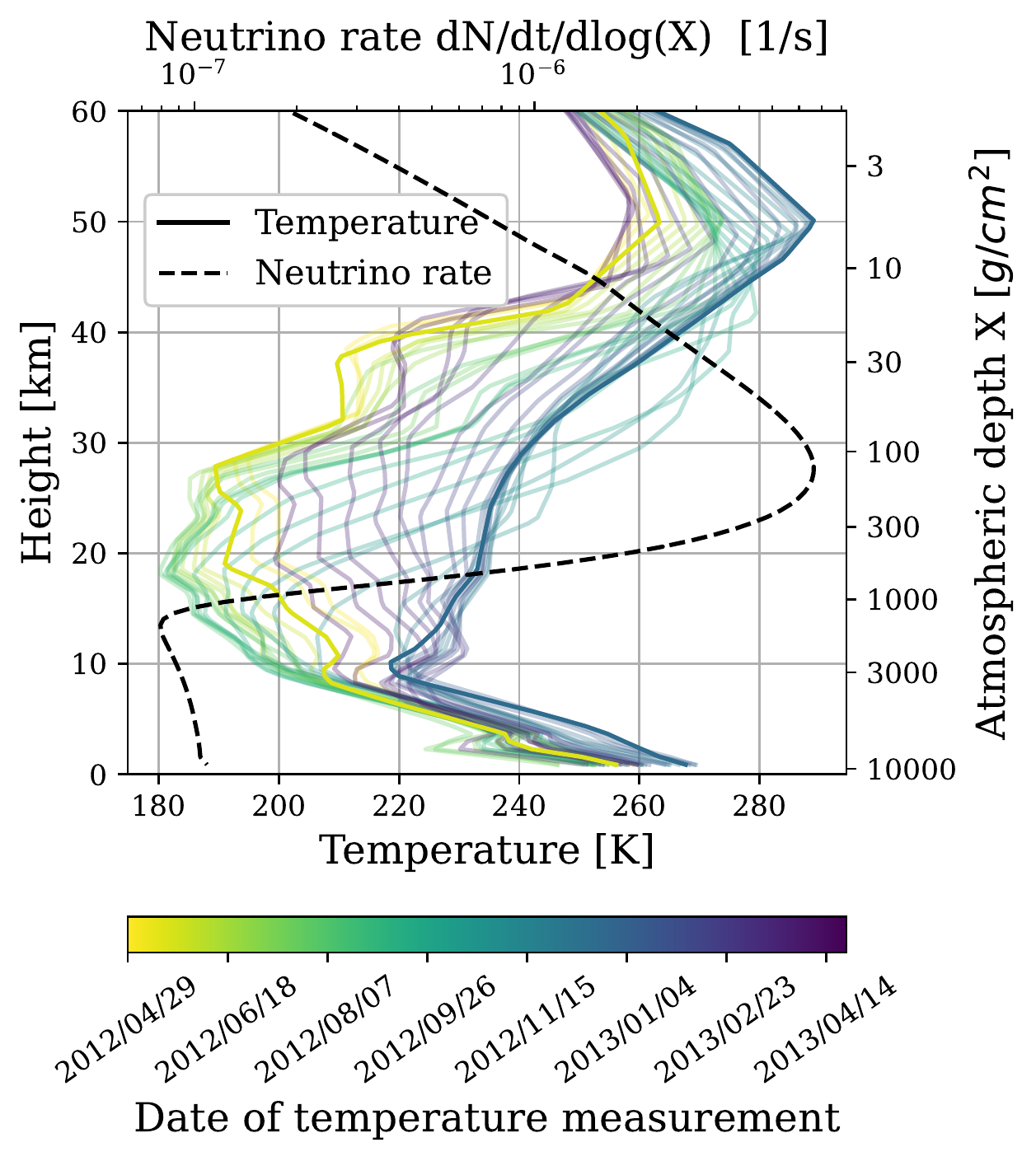}
    \caption{Example of altitude-profiles of atmospheric temperatures for an IceCube zenith of 95$^{\circ}$ throughout the years 2012 and 2013. The location on Earth corresponds to $-80^{\circ}$ in latitude and 0$^{\circ}$ in longitude. Also shown is the average effective production-profile of atmospheric $\nu_{\mu}$ in dashed black. These profiles are based on calculations with MCEq and are integrated in energy with the IceCube effective detection area. The small rise towards lower altitude is not related to meson decays but to muon decays in flight. }
    \label{fig:atmosphericSpectrum}
\end{figure}
As an example, 
Fig.  \ref{fig:atmosphericSpectrum} shows for one location the yearly variation of measured temperatures and the predicted effective neutrino production yield that is weighted with the detection efficiency of IceCube.
The figure highlights regions in the atmosphere with temperature variations that are  relevant for atmospheric $\nu_{\mu}$ detected by IceCube.


\subsection{The Effective Temperature
\label{sec:Teffective}}

The geographic location of the parent air shower of an observed atmospheric $\nu_{\mu}$ is given by the observed direction of the neutrino (see Fig.\ \ref{fig:NeutrinoPath}) that is expressed by the zenith angle $\theta $ and the azimuth angle $\varphi $.
The atmospheric altitude of production cannot be measured by IceCube, and thus the atmospheric temperature measured by AIRS has to be averaged along the line of sight $l$ as shown in  Fig.  \ref{fig:NeutrinoPath}.
The altitude $h$ of production is related to the 
 atmospheric depth
 $X(h,\theta,\varphi) = \int_{l}^{\infty}dl' \rho(l'(h,\theta,\varphi))  $ which integrates the density from the upper edge of the atmosphere along the line of sight towards the observer.
The calculated average is referred to as effective temperature:
\begin{equation}\label{eq:TeffDef}
    T_{\mathrm{eff}}(\theta,\varphi,t) = \frac{\int dX \cdot T(X,\theta,\varphi,t) R_X(\theta,\varphi, X, T) }{\int dX R_X(\theta, \varphi, X,T)}.
\end{equation}
Here, $t$ is the neutrino arrival time, $\theta$ and $\varphi $ are the local IceCube zenith and azimuth direction. These are, because of the location of IceCube at the geographic South pole, closely related to Earth's latitude and longitude. The depth dependent neutrino rate $R_X$ is defined by: 
\begin{equation}
    R_X(\theta, \varphi, X, T) = \int dE \; P(X,E,\theta,T(X, \theta, \varphi, t) ) \; A_{\mathrm{eff}}(E,\theta).
\end{equation}
Here $E$ is the neutrino energy and $P(X,E,\theta,T)$ is the atmospheric $\nu_{\mu}$ production yield, defined by $\int dX \; P = \frac{d\Phi}{dE}$ in~\cite{GaisserBook}, with $\Phi$ being the the atmospheric $\nu_{\mu}$ flux. $A_{\mathrm{eff}}(E,\theta)$ is IceCube's effective area, estimated by Monte-Carlo simulations. The temperature $T$  depends on the direction $\theta $ and $\varphi $  and the time $t$.
 The integral is approximated  by the  sum over the pressure levels given by AIRS, with the atmospheric depth being calculated for each level using the pressure, temperature and altitude. Both the neutrino production yield \linebreak
 $P(X,E,\theta,T)$ in the atmosphere and the effective 
detection area $A_{\mathrm{eff}}(E,\theta)  $ of IceCube depend on energy and zenith, and have to be integrated when
the atmospheric 
temperature is weighted with these quantities. The temperature data is discussed in  \ref{sup:tempdata}. A detailed description of the implementation is given in  \ref{sup:integrate}. 

In this analysis, we neglect the angular dependency of effective temperatures for the selected field of view and only consider time information for the correlation. Therefore, the angular dependence is averaged over the entire considered zenith range from \SIrange{90}{115}{\degree}.
\begin{equation} \label{eq:teffint}
    T_{\mathrm{eff}}(t) = \frac{\int d\Omega \; T_{\mathrm{eff}}(\theta,\varphi,t) \cdot \int dX R_X(\theta, \varphi, X,T)  }{\int d\Omega \; \int dX R_X(\theta, \varphi, X,T)   }.
\end{equation}
Here, $d\Omega = d\varphi \cdot d\cos(\theta)$ is the solid angle as seen from IceCube. The zenith angle $\theta $ can be approximately related with the geographic latitude $l$ by $\theta \approx l / 2 + \SI{135}{\degree}$ for $\theta>\SI{90}{\degree}$.  IceCube's azimuth $\varphi $ and Earth's longitude $\lambda$ are related by $\varphi = -\lambda + \SI{90}{\degree} $ if $\lambda \le \SI{90}{\degree}$ and $\varphi = -\lambda + \SI{90}{\degree} + \SI{360}{\degree}$ if $\lambda > \SI{90}{\degree}$. 
To correctly account for the differences of local measurement times with respect to the IceCube time zone (UTC), these respective local times are converted into UTC. Then, for averaging the global $T_{\mathrm{eff}} $,
 the temperatures are interpolated for all directions considered in the analysis
with respect to the UTC times of the temperature measurements and then averaged. More details are found in \cite{MasterThesisSimonHauser}.
 The resulting temperatures for each UTC day are  represented as the red line in Fig.  \ref{fig:daily_month_rate_temp}. The temperature data clearly follow the annual seasons and also the daily neutrino rates, shown in blue, indicate a  seasonal variation. The  monthly averaged neutrino rates in orange highlight the correlation, which is analyzed quantitatively in the next section.

\section{Analysis Methods and Results
\label{sec:results}}
This section introduces two methods to analyze the data: A binned $\chi^2$ fit, and an un-binned likelihood. The $\chi^2$-fit is used in similar atmospheric muon analyses, and gives a result which is comparable to \cite{Tilav:2019xmf}. The $\chi^2$-fit assumes gaussian distribution of event counts, which holds true for large event counts introducing a potential bias. To avoid dependencies on the statistical distribution of events an un-binned likelihood method is used as a complementary measurement. The un-binned likelihood is a new method, which is optimized for the smaller neutrino rates, and can be extended to account for an angular dependence in future analyses.
\subsection{Binned $\chi^2$-Fit
\label{sec:chisq}}

\begin{figure}
    \centering
    \includegraphics[width = 0.49\textwidth]{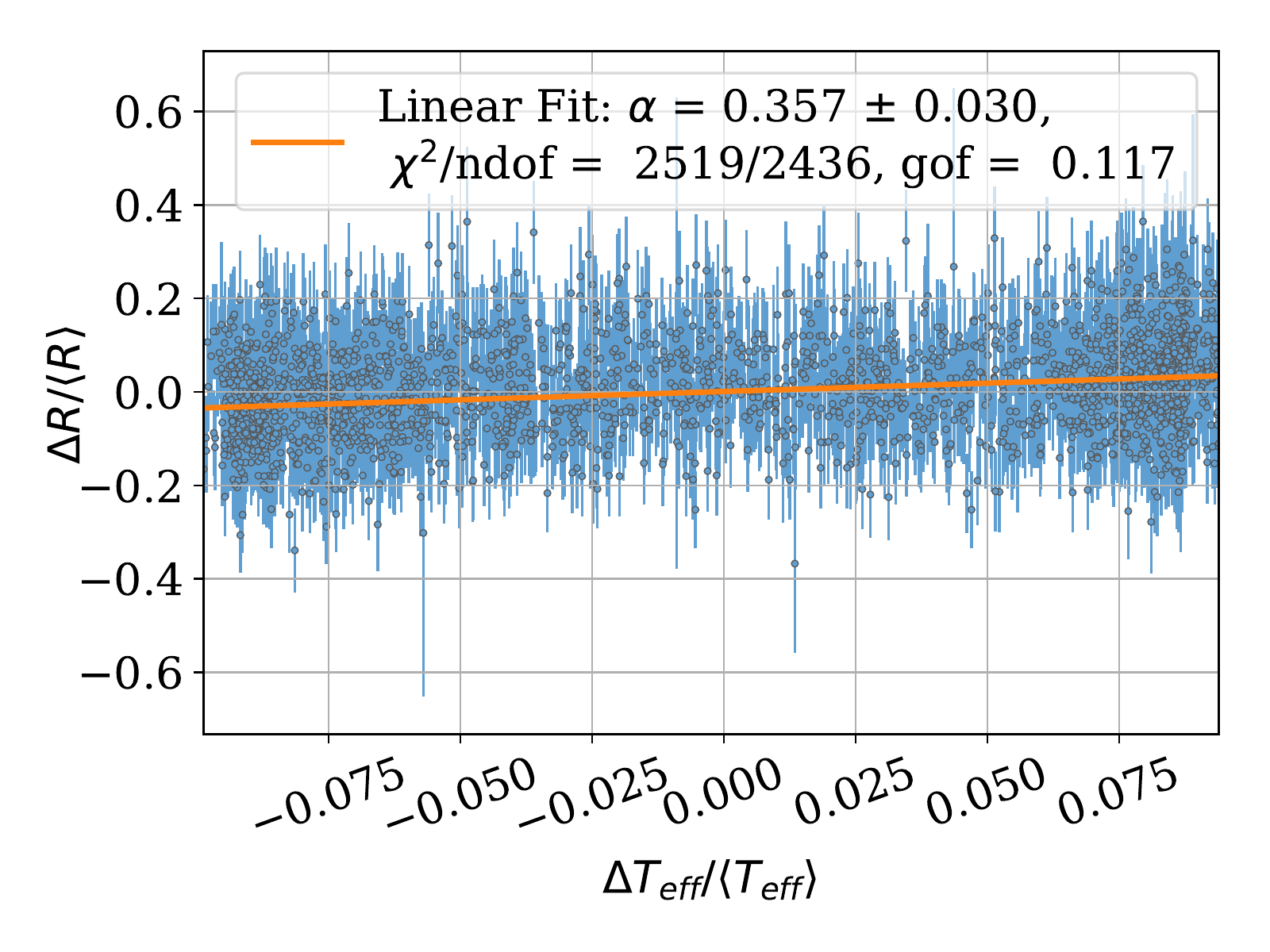}
    \caption{Plot showing the correlation of the measured relative rates of atmospheric $\nu_{\mu}$ and the relative effective temperature change. The orange line depicts the result of the $\chi^2$ fit of the shown data, with the values written in the legend. ``gof`` refers to the goodness of fit p-value, calculated from the $\chi^2$-value.}
    \label{fig:Chi2Fit}
\end{figure}

To estimate the correlation between the atmospheric temperatures and the measured atmospheric $\nu_{\mu}$ rates, a linear relation between the relative neutrino rate change and relative effective temperature change is approximated by \cite{GaisserBook,Heix:2019jib}
\begin{equation}
\label{eq:linregdef}
    \frac{R(t)-\bar{R}}{\bar{R}} = \alpha \frac{T_{\mathrm{eff}}(t)- \bar{T}_{\mathrm{eff}} }{\bar{T}_{\mathrm{eff}}},
\end{equation}
where $\alpha$ is the slope parameter,  $\bar{R}$ the average measured neutrino rate of the whole observation time,
and $ \bar{T}_{\mathrm{eff}}$ the corresponding average effective temperature. The parameter $\alpha$ measures the strength of the correlation.
The neutrino rate $R_i$ for each day $i$ is calculated by dividing the neutrino count $N_i$ by the effective detector livetime $\tau_{i}$ of that day. 
The effective temperature is evaluated at noon (GMT) of each day through interpolation. 
Large uncertainties that stem from the primary cosmic-ray flux and systematic detector effects cancel out by focusing on the relative change of rate and temperature.
To estimate $\alpha$, similar to muon seasonal variations analyses \cite{Tilav:2019xmf,Adamson:2014xga}, the $\chi^2$ is minimized:
\begin{equation}
    \chi^2 = \sum_i^{N_{\mathrm{tot}}} \frac{\Big(\frac{R_i-\bar{R}}{\bar{R}} - \alpha \frac{T_{\mathrm{eff}; i}- \bar{T}_{\mathrm{eff}} }{\bar{T}_{\mathrm{eff}}} - b \Big)^2  }{\sigma^2_{i}}.
\end{equation}
In this $\chi^2$-definition an additional bias parameter $b$ is added. This ensures robustness against systematic shifts in the average temperature and neutrino rate. The uncertainty is approximated by $\sigma_i = \sigma_{R_i}/\bar{R}$ and $\sigma_{R_i} = \sqrt{N_i}/\tau_{i}$. As the uncertainty is dominated by the limited amount of statistics of neutrino events, we neglect the uncertainty of the effective temperature and systematic uncertainties in the fit. 
The correlation of the quantities in Eq.~\ref{eq:linregdef} is plotted in Fig.~\ref{fig:Chi2Fit}, together with the result of the $\chi^2$ fit, with the best fit parameter $\alpha=0.357\pm0.030$. The p-value corresponding to a $\chi^2/ndof = 2519/2436$ is 0.117, which indicates agreement of the linear model with the data. The uncertainty of $\alpha $ is dominated by the amount of neutrinos measured per day, but also by the range of temperature variations which is limited to $\pm \SI{8}{\percent}$ (x-axis of Fig.~\ref{fig:Chi2Fit}). The observed bias is compatible with zero.
The significance using the $\chi^2$ difference between the observed $\alpha$ and the null hypothesis of $\alpha = 0$ is more than 10 standard deviations.
As discussed below in section~\ref{sec:systematics}, the resulting $\alpha $ has been tested for its robustness with respect to the used bin-size in time and it is found to be approximately unchanged, even if time-bins up to a month are used (see Fig.~\ref{fig:Binsize_dependence}). 

\subsection{Un-binned Likelihood}

\begin{figure}
    \centering
    \includegraphics[width = 0.45\textwidth]{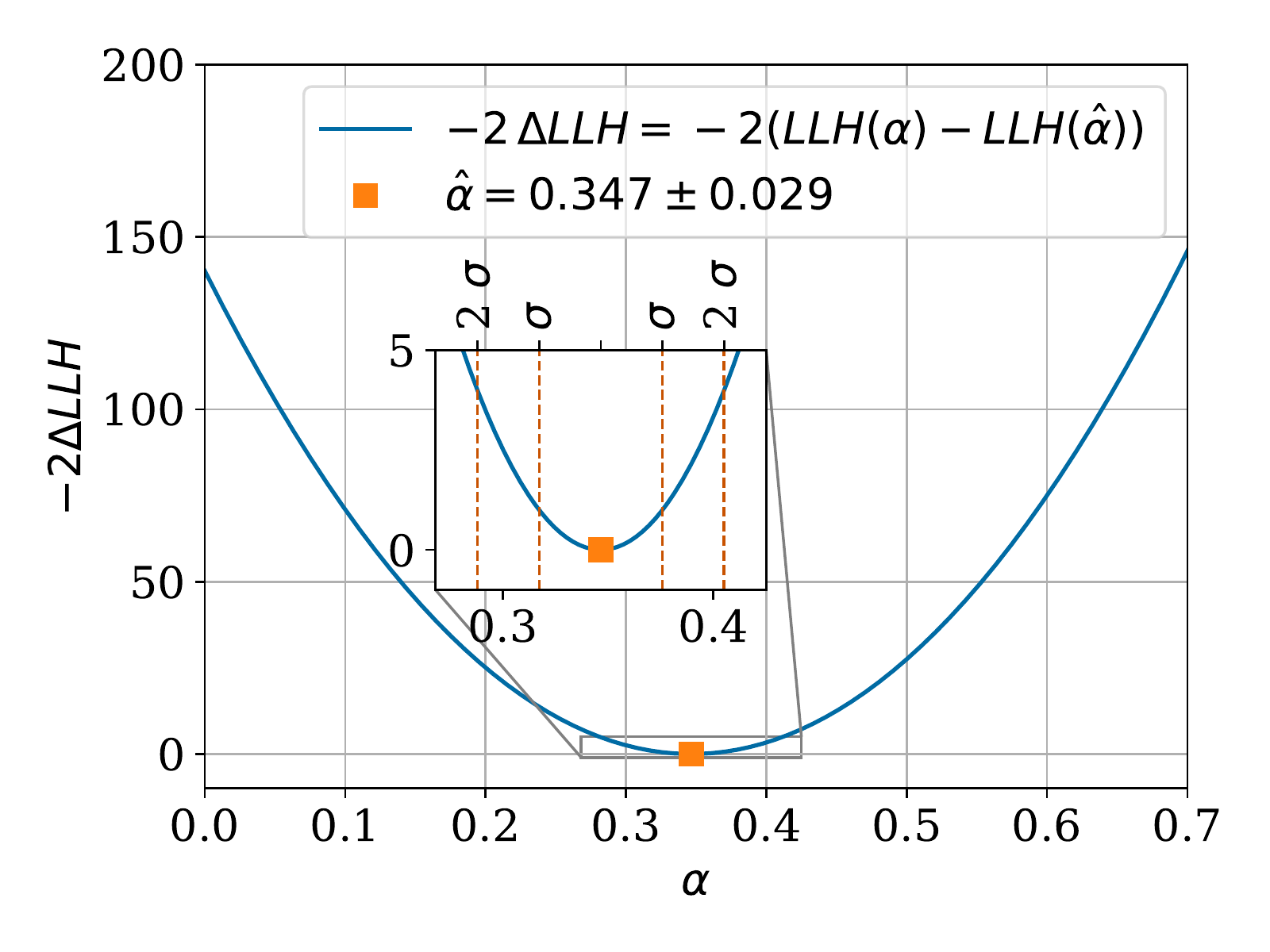}
    \caption{Figure showing the negative log-likelihood profile  for the un-binned likelihood defined in Eq.~\ref{eq:LLH}. In the inlay figure a zoom of the two sigma region is shown.}
    \label{fig:LLHFit}
\end{figure}

To exploit the full available time information of measured neutrinos, an un-binned likelihood technique is used. Compared to the $\chi^2$ fit this approach is also independent of the underlying statistical distribution of the events. This minimizes potential biases which could occur at small neutrino rates. In a linear approximation, the probability density distribution of neutrinos in time is given by 
\begin{equation}
    \label{eq:neutrinoprob}
    f(t) = \frac{1}{\tau_{\mathrm{tot}}} \Big(1 + \alpha \frac{T_{\mathrm{eff}}(t)- \bar{T}_{\mathrm{eff}} }{\bar{T}_{\mathrm{eff}}}\Big).
\end{equation}
Here, $\tau_{\mathrm{tot}}$ is the total effective livetime.
The total livetime is given by,
\begin{equation}
   \tau_{\mathrm{tot}} = \int_{\mathrm{obs}} dt \; \epsilon(t)
\end{equation}
with the measurement efficiency $\epsilon(t)$ defined as 1 if the detector is running and  0 if it is not taking data.
In the same way the average effective temperature is calculated
\begin{equation}
    \bar{T}_{\mathrm{eff}} = \frac{1}{\tau_{\mathrm{tot}}} \int_{\mathrm{obs}} dt \; \epsilon(t)\cdot T_{\mathrm{eff}}(t) ~.
\end{equation}
The log-likelihood is given by evaluating the probability density at the time of each measured neutrino event $i$  and then summing up:
\begin{equation}
\begin{split}
LLH & = \sum_i^{N_{\mathrm{tot}}} \log(f(t_i)) \\
    & = \sum_i^{N_{\mathrm{tot}}} \Big( \log(1 + \alpha \frac{T_{\mathrm{eff}}(t_i)- \bar{T}_{\mathrm{eff}} }{\bar{T}_{\mathrm{eff}}})-\log(\tau_{\mathrm{tot}})\Big).
\end{split}
\label{eq:LLH}
\end{equation}
Minimization of the negative $LLH$ with respect to $\alpha$ results in the best estimate for the temperature coefficient.
The resulting likelihood-profile in dependence of $\alpha$ can be seen in Fig.~\ref{fig:LLHFit}, together with the result of the minimization and the uncertainty estimated by the width of the parabola at $-2\cdot \Delta LLH = 1$: $\alpha=0.347\pm 0.029$. The uncertainty on $\alpha$ is comparable to the $\chi^2$ result. There is also a small shift in $\alpha$, which is consistent within the uncertainties of the two analysis methods.
The gain from the additional time information in the likelihood-method is found to be small. However, in a future analysis this un-binned likelihood approach can be extended to include also directional information without the need to average temperatures over the observation zone.

\subsection{Systematic Uncertainties \label{sec:systematics}}

For the discussion of systematic uncertainties, the contributions of the relative neutrino rate and  the relative effective temperature including the effect of averaging of data over the Southern hemisphere have to be considered. 

Uncertainties of the neutrino rates may arise from the estimation 
of the effective livetime and the possible contamination by background from wrongly reconstructed atmospheric muons, both of which are found negligible.
The atmospheric muon rate
 is correlated with the same seasonal phase as the neutrinos in the Southern hemisphere but with a correlation factor about twice as large \cite{Tilav:2019xmf}.
The contamination of the data sample with atmospheric muons has been estimated  in \cite{IceCube:2021uhz} to about \SI{0.15}{\percent}. This is further reduced by restricting the zenith angle to $\theta > \SI{90}{\degree}$, leading to a rate of less than \SI{0.16}{\per \day}.
 Assuming a maximum change of relative temperatures at the level of 8$\%$, the variation of the rate is negligible and amounts to less than $\Delta R \approx \SI{0.009}{\per \day}$. 
 As mentioned in section \ref{sec:IceCube}, IceCube is operating with a duty cycle close to \SI{97}{\percent}. The uncertainty of the \SI{1}{\percent} downtime is further reduced by excluding specifically days of low duty cycle from the analysis. 
 The remaining loss in livetime that could occur, for example during run transitions, can be calculated to an accuracy of a few microseconds.

Estimating the uncertainty of $T_{\mathrm{eff}}$ is more challenging.
The accuracy of the measured temperatures is estimated depending on the height \cite{AIRSReference} as \SI{1}{\kelvin \per \kilo \metre} (see section \ref{sec:AIRS}).  Assuming additionally a height uncertainty of 1\% in the integration of the atmospheric depth, this implies an uncertainty of $T_{\mathrm{eff}}$ of 0.25 \% or about \SI{0.7}{K}.  This estimation is supported by the observed difference of the estimated
$T_{\mathrm{eff}}$ between the ascending and descending measurements of each day for which a standard deviation of \SI{0.45}{K} is found.

\begin{figure}
    \centering
    \includegraphics[width=0.49\textwidth]{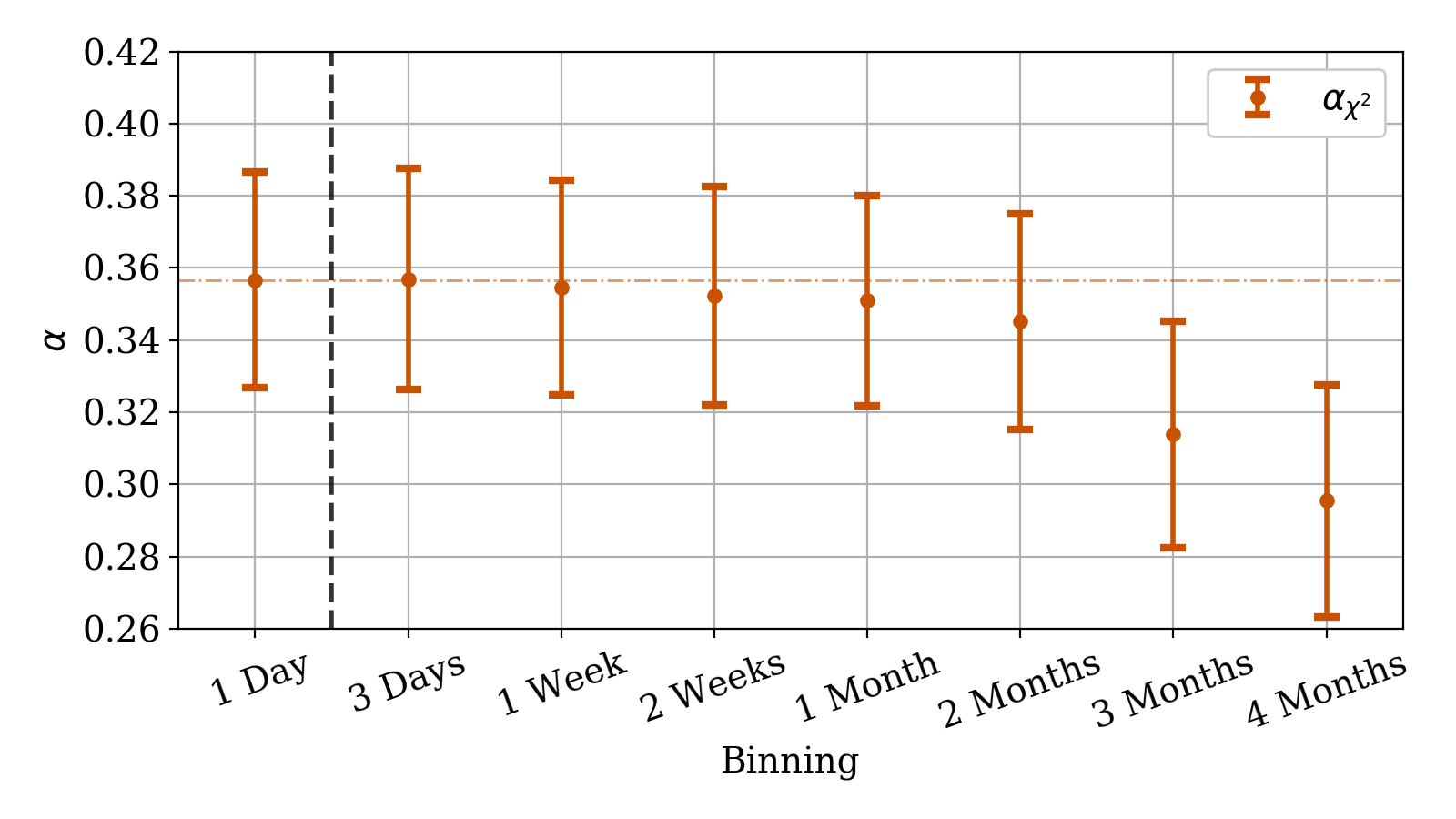}
    \caption{Plot showing the dependence of $\alpha$ on
the bin-size obtained from the $\chi^2$-fit. Only statistical uncertainties on $\alpha$ are given. }
    \label{fig:Binsize_dependence}
\end{figure}

The sufficient coverage of the atmosphere in time by the satellite can be tested with the binned analysis by increasing the bins from days to longer periods and using only the temperature at the mean time of the bin.
The result is shown in Fig.  \ref{fig:Binsize_dependence}. Even if increased to the scale of a month, the obtained $\alpha $ value remains almost unchanged. Only on time-scales of 3 months or more does the correlation decrease significantly due to the averaging of the seasonal effects.
This, in turn, is a strong indication that an even better coverage in time and correspondingly smaller time bins would result in an unchanged $\alpha$ and that the time coverage of the temperature data is sufficient.

As discussed above,  fractional ground-coverage and topographic changes ($< \SI{0.1}{\percent}$) can be neglected for the chosen Southern hemisphere region. 
It turns out that the largest uncertainty arises from the numerical integration of the slant depth. 
Depending on the choice of bin-boundaries in the atmospheric depth integration, the relative temperatures can change up to \SI{1}{\percent}. 
Another concern is the absolute accuracy of the temperature data. For estimating this uncertainty we have estimated the $T_{\mathrm{eff}}$ values with a partly independent data set of atmospheric temperature profiles that also include higher altitudes \cite{ERA5data}, see also  \ref{sup:tempdata}. 
The difference of the daily determined $T_{\mathrm{eff}} $ values has a standard deviation of  \SI{0.8}{\percent} which supports a robust estimation of $T_{\mathrm{eff}} $.

The median uncertainty of the directional reconstruction of neutrino events is roughly \SI{1}{\degree} at \SI{1}{TeV}  neutrino energy and improves to 
 \SI{0.3}{\degree} at \SI{1}{PeV}
\cite{IceCube:2018ndw}. This results in a small mismatch of the assumed location of the parent air shower from its true location. By averaging the $T_{\mathrm{eff}} $ over zenith and azimuth in the observation zone, this effect becomes negligible. Also other systematic detector effects, like the photo detection efficiency of the DOMs or optical properties of the ice, can be largely excluded. The detector is being operated in an almost unchanged configuration during the observation time and small changes in the effective area do cancel in the relative rates. 
Related to the detection of neutrinos with IceCube, additional uncertainties arise. This is tested by varying systematic effects of the detector, e.g.\  the absorption length in the ice or the quantum efficiency of the PMTs, in the MC simulation. The largest deviation of $\alpha$ was found to be $\delta \alpha = 0.005$, so the systematic detector effects can be neglected compared to other uncertainties.
Effects from the uncertainty of the primary cosmic-ray spectrum on the prediction are tested in two ways: The first method varies the spectral index of the primary spectrum by multiplying the flux with a factor $(E/E_0)^{\Delta \gamma_{CR}}$. A change of $\Delta \gamma_{CR}$ of  ~0.05 results in a change $\delta \alpha = 0.005$.
The second method changes the composition of the primary flux by fixing it to either pure proton ($\ln(A) = 0$) or pure iron ($\ln(A)=4$). The difference in $\alpha$ between these extreme scenarios is $~0.02$. 


\section{Comparison with Model Predictions
\label{sec:methods}}

In this section the experimental measurement is compared to model expectations that are based on the physics  of cosmic-ray air showers.

\subsection{Atmospheric $\nu_{\mu}$ Flux Modeling
\label{sec:MCEqPred}}

The model expectation of the seasonal variation of the neutrino flux is based
on a full numerical calculation of 
the cascade equation of atmospheric air showers using 
the numerical cascade equation solver MCEq~\cite{MCEqpaper}.
The calculation of the neutrino flux is done for each day at each location. Here we explicitly use the daily-measured local temperature profiles measured by AIRS, as they are used in the calculation of effective temperatures. Therefore, the specific properties of the local and time dependent atmosphere is accounted for as precisely as possible.
This approach thus includes the full seasonal variation of temperature profiles of the global atmosphere during  the observation time.
We also compare  the results of the analysis  to the expectation of an analytic approximation of the cascade equation~\cite{GaisserBook}, which is also used to estimate the production probability in Eq. \ref{eq:TeffDef}. The analytic approximation embeds the temperature dependence in the critical energies $\epsilon_{\pi,K}$. The critical energy is the energy scale above which re-interactions dominate decay processes of the parent mesons. Both of these approaches are compared in \cite{MCEqAAComparison}.
For the calculations with MCEq, hadronic interactions are modeled by 
SIBYLL~2.3c \cite{SIBYLLPaper} and the cosmic-ray primary particle flux by the H4a flux model~\cite{H4aPaper}. 
After integrating the modeled neutrino fluxes $\Phi(E,\theta,\varphi,t)$ with the effective area of IceCube $A_{\mathrm{eff}}(E,\theta)$, this method gives the expected neutrino rates $R(t)$ for each day during the observation time:
\begin{equation}
    R(t) = \int dE \; d\Omega \; A_{\mathrm{eff}}(E,\theta) \frac{d\Phi(E,\theta,\varphi,t)}{dE}.
\end{equation}
From this time dependent rate expectation, we generate an ensemble of pseudo-experiments of neutrino events for the considered observation time that are then analyzed using the same methods and the same effective temperature data as the experimentally measured neutrino events. 

A comparison of the expected correlations with the experimental results for the two analysis methods, are shown in Table \ref{tab:results}. In addition to the full MCEq calculation and the analytic approximation, the expectation from the extreme cases of only pions or kaons as parent particles are given. The agreement between the analytic approximation and the MCEq based result is high, underlining the result in \cite{MCEqAAComparison}.
 The experimental measurement finds a correlation that is smaller than the theoretical prediction by about \numrange{2}{3} standard deviations. 
 This tension is investigated further in the next sections. 

\begin{table}[ht]

\caption{Table of the experimental results and predictions for $\alpha$ using the binned $\chi^2$ and un-binned LLH analysis methods. The prediction for MCEq~\cite{MCEqpaper} used the SIBYLL 2.3c~\cite{SIBYLLPaper} hadronic interaction model. The kaon and pion results rely on the flux fraction of MCEq to estimate their neutrino production rate. The uncertainty given for the predicted values corresponds to the expected statistical uncertainty for each measurement.}
\begin{tabular}{l|c|c}
\label{tab:results}
$\alpha$       & LLH             & $\chi^2$        \\ 
\hline
Exp. result    & 0.347$\pm$0.029 & 0.357$\pm$0.030 \\ 
MCEq           & 0.424$\pm$0.038 & 0.424$\pm$0.039 \\ 
Analytic Appr. & 0.429$\pm$0.038 & 0.439$\pm$0.039 \\ 
Kaons only         & 0.278$\pm$0.076 &                 \\ 
Pions only         & 0.637$\pm$0.099 &                 \\ 

\end{tabular}
\end{table}

\subsection{Uncertainties of the Atmospheric $\nu_{\mu}$ Flux}

\begin{figure}
  \includegraphics[width=0.49\textwidth]{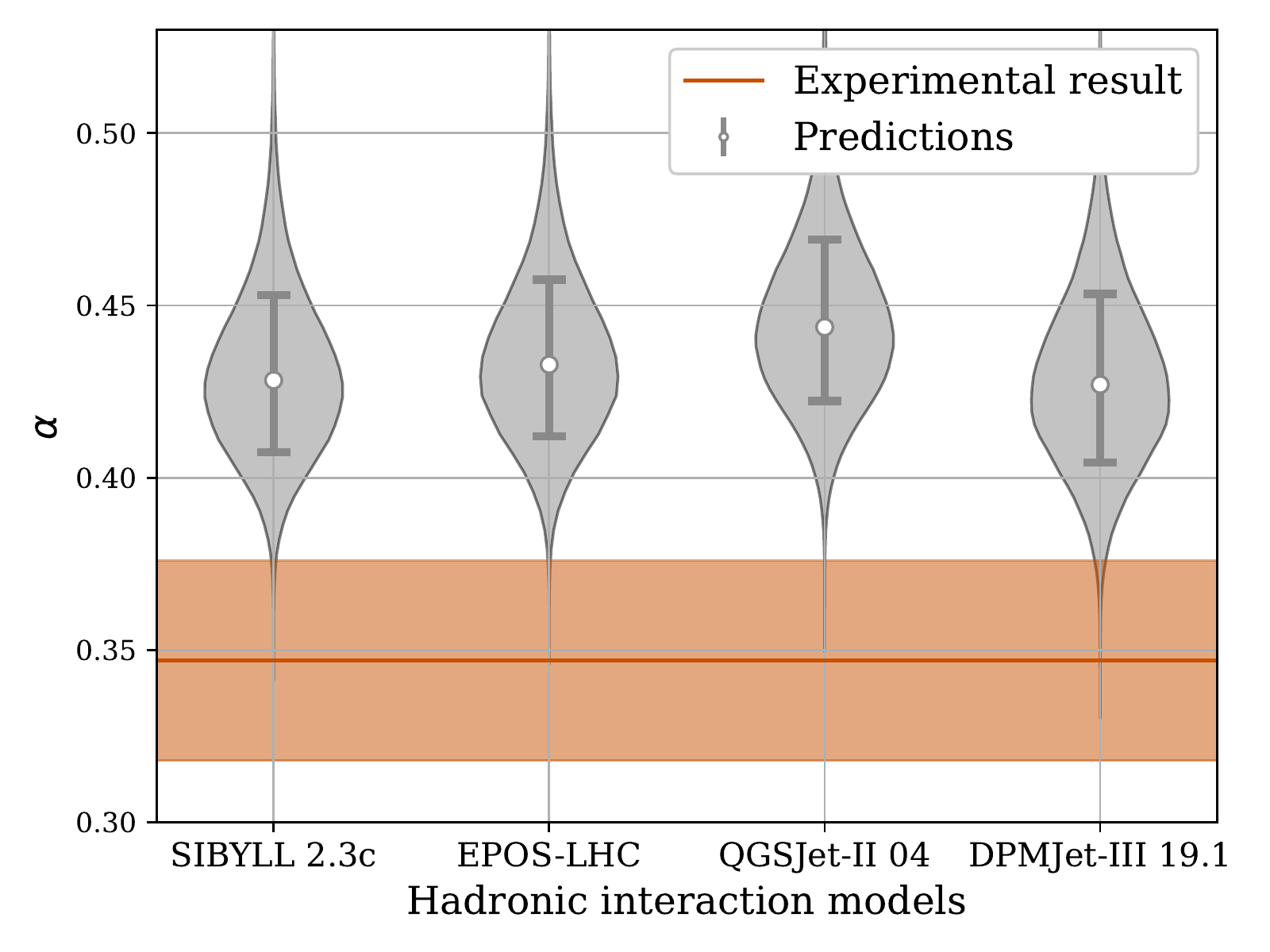}
\caption{Plot comparing the predictions (error-bar and violins) of different hadronic interaction models with the experimental result (horizontal red line) of this analysis (un-binned LLH). The systematic uncertainties of the predictions are estimated by varying several Barr parameters~\cite{BarrPaper} inside their prior distributions, with the resulting likelihood distribution of expected $\alpha $ values being represented by the violins (larger width corresponds to a higher likelihood). The p-values of each hadronic interaction model are estimated in Table \ref{tab:pvalues}.}
\label{fig:Hadronic_models}       
\end{figure}

Uncertainties in the prediction of the flux 
 of atmospheric muon neutrinos are related to the flux of primary cosmic-rays, the composition of the primary particles, and the production and re-interaction probability of parent mesons (ka\-ons and pions) in the atmosphere. In particular, a substantially larger number of parent kaons (see Table \ref{tab:results})
 would reduce the observed tension.
 Changes in the fraction of parent mesons can be related  to uncertainties in the respective production and re-interaction cross sections of these mesons in  particle interactions in the air. Such uncertainties are described in~\cite{BarrPaper}, by introducing parameters in different regions of primary nucleon energy and secondary meson lab energy fraction $x_L$. 
 As shown in \cite{IceCube:2021uhz}, only high-energy ($>\SI{100}{ \;GeV}$) meson production is relevant for the atmospheric $\nu_{\mu}$ flux. 
 For an estimation how these uncertainties propagate to the expected seasonal variation of atmospheric $\nu_{\mu}$ fluxes, we have varied these parameters assuming independent Gaussian uncertainties for each of these and calculated the corresponding value of $\alpha $ for each instance. This procedure is repeated, in addition to the SIBYLL 2.3c model \cite{SIBYLLPaper}, also for different hadronic interaction models: EPOS-LHC \cite{EPOSLHCPaper}, QGSJet-II 04 \cite{QGSJetPaper},
 and DPMJet-III 19.1 \cite{DPMJetPaper}.
 The resulting distributions of $\alpha $ values are displayed in Fig.  \ref{fig:Hadronic_models}.
It can be seen that all interaction models predict rather similar $\alpha$ values with also a similar variance when including the variation of the parameters in \cite{BarrPaper}. The observed tension thus persists also under variation of the atmospheric and hadronic model parameters.

For each of the hadronic interaction models, a p-value for the agreement with the experimental result based on the statistical uncertainty of the experiment and the systematic uncertainty from the atmospheric modeling is calculated and shown in Table \ref{tab:pvalues}. The best agreement is found for the DPMJet-III 19.1 model, with a tension of 1.8 $\sigma$, the largest tension is found for the QGSJet-II 04 model with a tension of 2.3 $\sigma$. 

\begin{table}[b]

\caption{Summary of the expected $\alpha$-values for different hadronic interaction models with the uncertainty estimates based on the method in~\cite{BarrPaper}. The given uncertainties reflect the 68\% region around the central value. The p-value giving the comparability of prediction and experimental result includes the statistical uncertainty of the experiment as well as the uncertainty estimated by varying the Barr-parameters. All predictions used MCEq with the H4a primary cosmic ray model.}
\begin{tabular}{l|c|c}
\label{tab:pvalues}
hadr. interaction model & $\alpha$                  & p-value \\ \hline
SIBYLL 2.3c~\cite{SIBYLLPaper}     & 0.429$_{-0.021}^{+0.025}$ & 0.025   \\ 
EPOS-LHC~\cite{EPOSLHCPaper}         & 0.433$_{-0.021}^{+0.025}$ & 0.019   \\ 
QGSJet-II 04~\cite{QGSJetPaper}   & 0.443$_{-0.022}^{+0.025}$ & 0.010   \\ 
DPMJet-III 19.1~\cite{DPMJetPaper}  & 0.426$_{-0.023}^{+0.026}$ & 0.033   \\ 
\end{tabular}
\end{table}

\section{Systematic Effects of the Observed Tension
\label{sec:systematics_splits}}

\begin{figure}[t]
    \centering
\includegraphics[width = 0.49\textwidth]{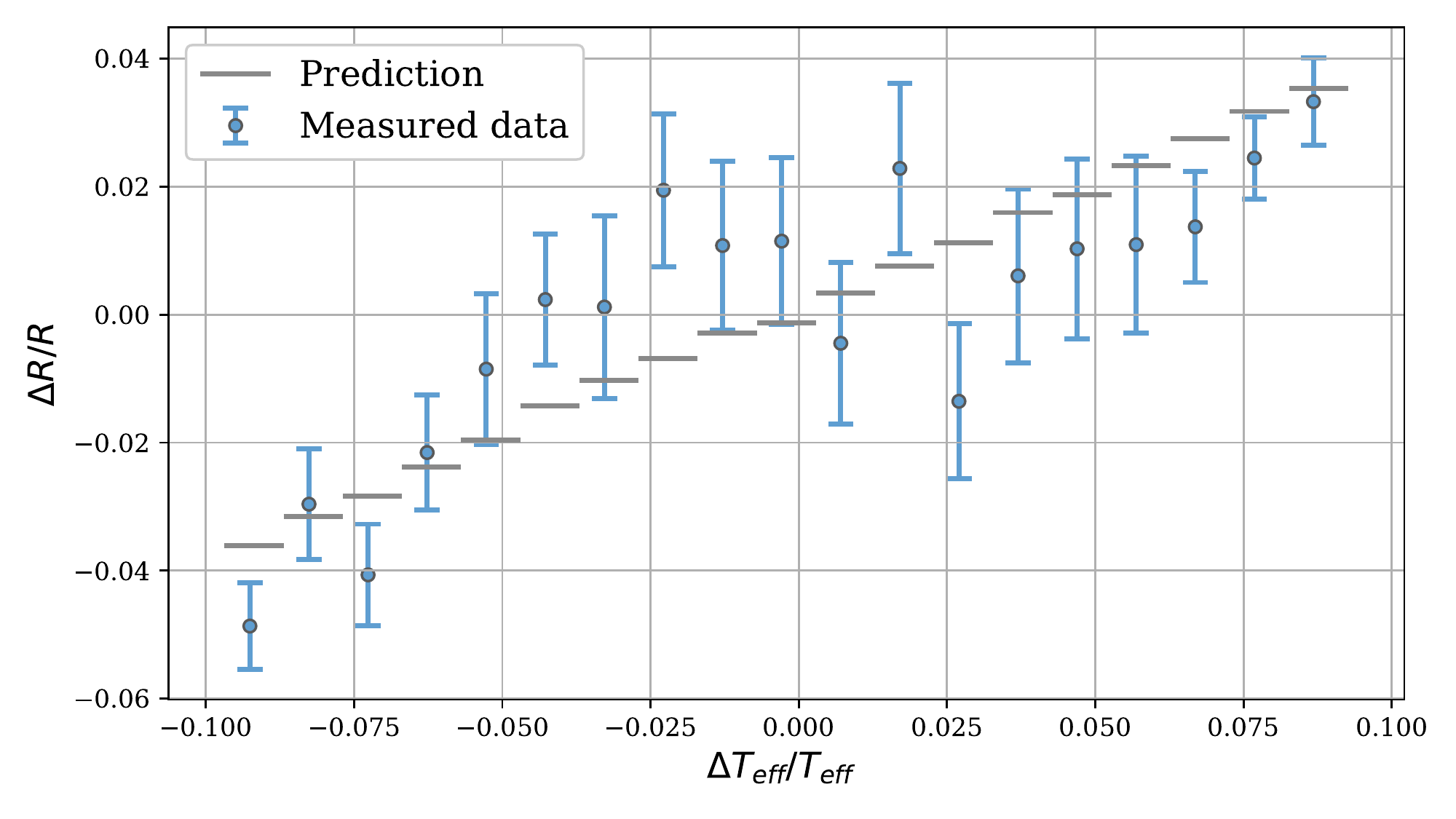}
    \caption{Plot depicting the predicted and measured relative neutrino rate against the relative temperature variations bins. The slope corresponds to $\alpha$. The prediction is made using
MCEq with the SIBYLL 2.3c hadronic interaction model.}
    \label{fig:Tempbins}
\end{figure}

In order to investigate the origin of the observed tension between model predictions and our measurement, Fig.  \ref{fig:Tempbins} shows the measured and expected rate variation as a function of the respective temperature difference.
The figure indicates an inverted sigmoidal trend of the data for small values of $\Delta T_{\mathrm{eff}}$.
While 
the extreme values of $\Delta T_{\mathrm{eff}} $ 
would be consistent with the expectation, the overall range results in a smaller overall slope for the experimental data.
The apparent region close to small absolute $\Delta T_{\mathrm{eff}}$ corresponds to the spring and fall seasons at the South Pole. These times are dominated by  rapid temperature changes with large fluctuation particularly in spring as seen in Fig.  \ref{fig:daily_month_rate_temp}.


\begin{figure*}
    \centering
    \includegraphics[width=1.0\textwidth]{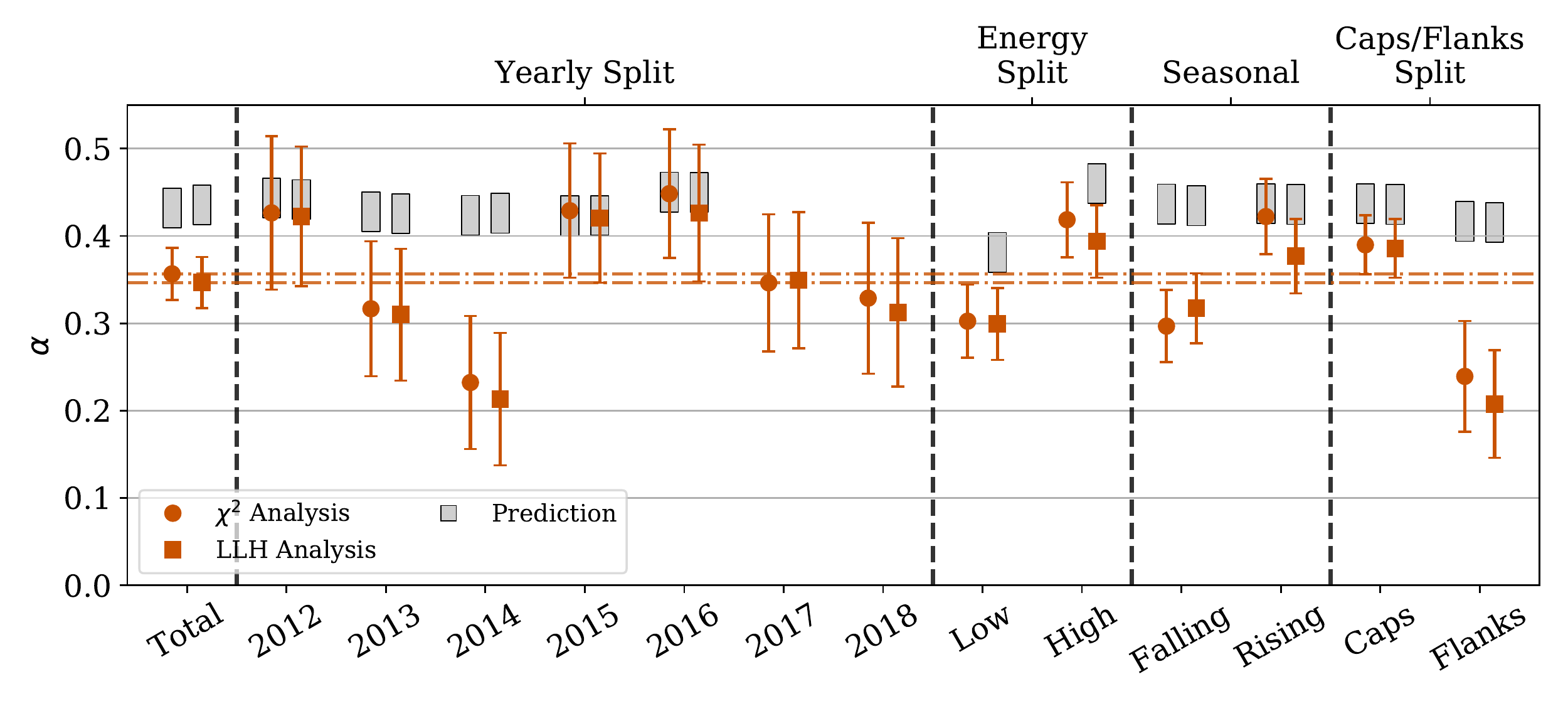}
    \caption{Plot summarizing the main results for the slope parameter $\alpha$, after applying either the binned $\chi^2$ fit (dots) or the un-binned likelihood analysis (squares). This is done for the full data set (red, first column), for each individual year starting in May (red, second column) and three different splits of the data: By energy (third column), by rising and falling temperatures (fourth column) and by splitting into caps and flanks (last column). For comparison, predictions of the respective data are shown as gray bands, with uncertainties being estimated using the approach described in~\cite{BarrPaper}. The dashed red line is an extension of the overall result for $\alpha$, to compare it to the individual results.}
    \label{fig:Summary_plot}
\end{figure*}

In the following sections, we investigate the effects of splitting the data  into smaller, systematically split sets. The results of the different splits are summarized in Fig.  
\ref{fig:Summary_plot}.

\subsection{Yearly Splits}

When splitting the observation periods into single years and repeating the analysis, 
the obtained $\alpha $ values show larger fluctuations.
These are, however, consistent with the larger statistical uncertainties.
For the years 2012, 2015, and 2016 the obtained values agree with the expectation, while for the year 2014 the smallest $\alpha $ value is measured.
Detailed investigations of specific peculiarities of the IceCube detector operation as well as the used AIRS temperature data have revealed no indications of any specific difference for this period.
Therefore, the observed fluctuations away from the all year mean are considered of statistical origin.


\subsection{Energy Dependence}

The correlation between the atmospheric $\nu_{\mu}$ rate and the atmospheric temperature exhibits a strong energy dependence, and up to this point the correlation analysis effectively averages the effect over the  energy distribution of measured neutrinos. In order to verify the expected energy dependence and the implicit averaging of the spectrum, the full data is split into two samples with respect to low and high reconstructed energy~\cite{IceCubeEreco}.
The measured energy distribution is shown in Fig.  \ref{fig:splits_description}, and the data is split at the median reconstructed energy value of \SI{700}{GeV} which thus yields two samples of equal statistical power.
The analysis is repeated for each sample separately and results are shown in 
Fig.  \ref{fig:Summary_plot}, together with the theoretically predicted values for the same energy split. 
As expected, in the low-energy bin the value of $\alpha$ becomes smaller in contrast to the larger value at high energies. The same effect is observed for the prediction as well. This reflects that the decay probability decreases compared to the interaction probability of the parent mesons with higher energy, leading to a larger seasonal dependence. 
 
The tension between model and experiment is similar for both energy regions. This observation disfavors  effects related to the energy dependence of the detector response or the prediction
as origin of the observed tension. One can conclude that the tension persists independent of the selected energy range.

\subsection{Hysteresis due to Seasonally Dependent Atmospheric Layering.}

During seasons of generally rising or generally falling temperatures, the layering of  the temperatures can strongly differ (see Fig.  \ref{fig:atmosphericSpectrum}).
Due to the marginalization of the altitude  information in the calculation of $T_{\mathrm{eff}} $, different temperature profiles in spring and fall can result in a small difference of the  atmospheric $\nu_{\mu}$ rates for the same value of $T_{\mathrm{eff}} $.
This effect has been observed for atmospheric
 muons~\cite{Tilav:2019xmf}, and is called 
  hysteresis effect because of the difference in rate for different seasons at the same $T_{\mathrm{eff}} $ value.
  The non-linearity in the relation between rate and temperature is expected on the level of less than  \SI{1}{\percent} difference of the measured rates between spring and fall which is much smaller than the effect observed here.
  Furthermore, our calculation of the expectation that includes each measured temperature profile  does in fact include this effect
   and results in a small expected difference between spring and fall of less than \SI{0.4}{\percent}.

 As an experimental verification,  the data is split at the maximum and minimum points of temperature into a falling (fall) and rising (spring) samples (see Fig.  \ref{fig:splits_description}). The results of the split is shown in Fig.  \ref{fig:Summary_plot}.
 A difference between the two split samples is seen
 with a larger value of $\alpha $ in the spring season, which however both are statistically compatible with the average value. 
 Unlike the experimental observation this large seasonal difference is not predicted by the theoretical calculation, despite having included the seasonal hysteresis effect into the calculation.
 
\subsection{Extreme Temperature Bins}
Following up the observed deviations in the spring and fall seasons, as seen in Fig.~\ref{fig:Tempbins} in the mid-high and mid-low relative temperature variations, we further investigate systematic effects in the observed data.
We separate the data into “caps” and “flanks” (see Fig.  \ref{fig:splits_description}). 
Here, caps correspond to seasons of extreme temperatures, i.e.\ winter and summer, and flanks to the data in the transition seasons.
To the right in Figure \ref{fig:Summary_plot}, \SI{25}{\percent} is included in each cap, i.e. \SI{50}{\percent} of all the data, and thus \SI{50}{\percent} of all the data in the flanks data set. As the lever arm of the extreme points in the fit remains the same compared to the complete data set, the uncertainty of the “caps” is smaller compared to the flanks. Again, the prediction has been calculated separately for the subsets of data being analyzed.
When limiting the analyses to the caps, the value of $\alpha $ is consistent with the theoretical expectation. On the other hand, the observed value of $\alpha$ of the flanks strongly disagrees with the predictions which do not depend on the chosen selection. The systematic shift observed in this split is similar to the shape in Fig. \ref{fig:Tempbins}. 
Further systematic studies based on data splits can be found in~\cite{MasterThesisHerpenbeck}. This also includes a zenith-dependent analysis of $\alpha$, which was excluded as it did not contain additional insights.

\section{Conclusion and Outlook
\label{sec:Conclusion}}

In this paper we present 
the analysis of the  correlation of the rate of high-energy atmospheric $\nu_{\mu}$  measured by IceCube with the effective atmospheric temperature based on atmospheric profiles measured by the AIRS instrument on the Aqua satellite.
For an observed 10\% variation of the effective atmospheric temperature we observe a 3.5(3)\% highly significant ($> 10 \; \sigma$) seasonal variation of the rate of high-energy atmospheric neutrinos.

In the comparison with the expectation from cosmic-ray air shower models predicting a larger seasonal variation of 4.3\%, a tension of about two to three standard deviations is observed. This tension
is marginally consistent with a statistical fluctuation, and
cannot be explained by known systematic uncertainties: neither the experimental measurement, nor the used satellite data, nor the modeling of air showers. In comparison to the muon seasonal variation analysis~\cite{bouchta1999seasonal,MinosSeasonalPaper,IceCubeSeasonalpaper}, we observe a smaller value for $\alpha$, which is expected due to the larger kaon contribution to the neutrino flux. The re-analysis of systematically selected subsets of the data shows deviations from the average model, with some being predicted (reconstructed energy) and others not being predicted (rising/falling temperatures, extreme temperatures). This is interpreted as a hint that the production of atmospheric $\nu_{\mu}$ during rapidly changing atmospheric conditions may not yet be fully understood.

The observed tension demonstrates that in the future  measurements of seasonal  variations of atmospheric $\nu_{\mu}$ may provide a new and complementary test of our understanding of the physics of atmospheric air showers.
As the present analysis is still limited by statistical uncertainties, a future analysis needs to include
a larger statistics data set. Five additional years of IceCube data taking should become available for a follow-up analysis soon.
In addition, the analysis itself can be expanded by
taking into account the atmospheric profile in the specific direction of the observed neutrino events. 
A substantially larger detector, IceCube-Gen2, \cite{IceCube:2014gqr} will allow observing the atmospheric temperature correlation with much increased statistics and thus allow for testing the correlation during shorter periods of time.

\begin{acknowledgements}
The IceCube collaboration acknowledges the significant contributions to this manuscript from Jakob Böttcher, Hannah Erpenbeck, and Christopher Wiebusch.We sincerely thank Christian von Savigny for valuable discussions and providing help on using the ECMWF data. We dedicate this publication to Tom Gaisser who laid the foundations of this analysis. We acknowledge the support from the following agencies:

 USA {\textendash} U.S. National Science Foundation-Office of Polar Programs,
U.S. National Science Foundation-Physics Division,
U.S. National Science Foundation-EPSCoR,
Wisconsin Alumni Research Foundation,
Center for High Throughput Computing (CHTC) at the University of Wisconsin{\textendash}Madison,
Open Science Grid (OSG),
Advanced Cyberinfrastructure Coordination Ecosystem: Services {\&} Support (ACCESS),
Frontera computing project at the Texas Advanced Computing Center,
U.S. Department of Energy-National Energy Research Scientific Computing Center,
Particle astrophysics research computing center at the University of Maryland,
Institute for Cyber-Enabled Research at Michigan State University,
and Astroparticle physics computational facility at Marquette University;
Belgium {\textendash} Funds for Scientific Research (FRS-FNRS and FWO),
FWO Odysseus and Big Science programmes,
and Belgian Federal Science Policy Office (Belspo);
Germany {\textendash} Bundesministerium f{\"u}r Bildung und Forschung (BMBF),
Deut\-sche Forschungsgemeinschaft (DFG),
Helmholtz Alliance for Astroparticle Physics (HAP),
Initiative and Networking Fund of the Helmholtz Association,
Deutsches Elektronen Synchrotron (DESY),
and High Performance Computing cluster of the RWTH Aachen;
Sweden {\textendash} Swedish Research Council,
Swedish Polar Research Secretariat,
Swedish National Infrastructure for Computing (SNIC),
and Knut and Alice Wallenberg Foundation;
European Union {\textendash} EGI Advanced Computing for research;
Australia {\textendash} Australian Research Council;
Canada {\textendash} Natural Sciences and Engineering Research Council of Canada,
Calcul Qu{\'e}bec, Compute Ontario, Canada Foundation for Innovation, WestGrid, and Compute Canada;
Denmark {\textendash} Villum Fonden, Carlsberg Foundation, and European Commission;
New Zealand {\textendash} Marsden Fund;
Japan {\textendash} Japan Society for Promotion of Science (JSPS)
and Institute for Global Prominent Research (IGPR) of Chiba University;
Korea {\textendash} National Research Foundation of Korea (NRF);
Switzerland {\textendash} Swiss National Science Foundation (SNSF);
United Kingdom {\textendash} Department of Physics, University of Oxford.
\end{acknowledgements}


\begin{appendix}

\section{Supplementary Material}

\subsection{Atmospheric Temperature Data \label{sup:tempdata}}

\begin{table}
\caption{Pressure levels used by AIRS measurement for temperatures and heights.}
\label{tab:presslevel}
{
\begin{tabular}{l l|l l| l l|l l}

$j$ & $p$  & $j$ & $p$  & $j$ & $p$  & $j$ & $p$ \\  & {[}hPa{]} &  & {[}hPa{]} &  & {[}hPa{]} &  & {[}hPa{]} \\
\hline 
1     & 1000        & 7     & 400         & 13    & 70          & 19    & 7           \\ 
2     & 925         & 8     & 300         & 14    & 50          & 20    & 5           \\ 
3     & 850         & 9     & 250         & 15    & 30          & 21    & 3           \\ 
4     & 700         & 10    & 200         & 16    & 20          & 22    & 2           \\ 
5     & 600         & 11    & 150         & 17    & 15          & 23    & 1.5         \\ 
6     & 500         & 12    & 100         & 18    & 10          & 24    & 1           \\ 
\end{tabular}
}

\end{table}

\begin{figure}
    \centering 
    \includegraphics[width =0.45\textwidth]{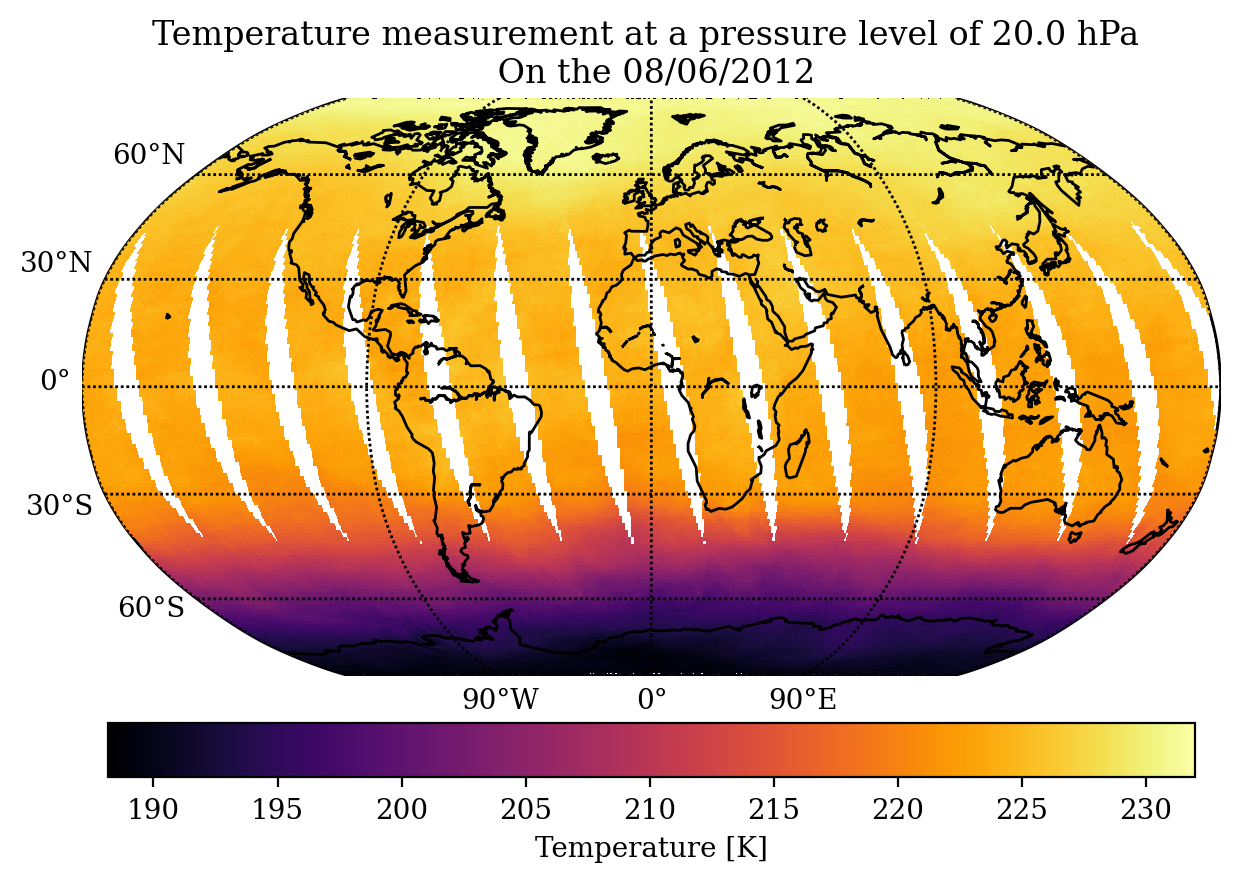}
    \caption{Plot of the temperature data taken on the 08/06/2012 of the \SI{20}{\hecto \pascal} pressure level. Missing data is colored in white. The gaps due to the limited swath are visible.}
    \label{fig:TempDataTaking}
\end{figure}

The data taken by AIRS is evaluated on 24 pressure levels given in Table \ref{tab:presslevel}. Due to the limited swath of the instrument, gaps in the coverage
of the available temperature data appear between two consecutive orbits that appear as longitudinal slices, as seen in the example of Fig.  \ref{fig:TempDataTaking}.
These gaps are largest at the equator but towards larger Northern and Southern latitudes they disappear because 
of better overlap of the swath of consecutive orbits.
The relevant zenith range of \SIrange{90}{115}{^\circ} for this analysis corresponds to geographical latitudes of the parent air shower from \SIrange{-90}{-40}{^\circ}.
At these latitudes gaps have  largely  disappeared (see Fig.  \ref{fig:TempDataTaking}).
Remaining gaps are interpolated with values close in time (prior and next day) as well as between neighboring longitudes. 
For high pressure levels close to sea level, the data is  limited by ground structure like mountains.
This is particularly the case for Antarctica. Due to the altitude of the continent and cold temperatures, no data is available for levels of high pressure.
As cosmic ray air showers are quickly stopped when reaching the ground,
and  
the production yield of atmospheric $\nu_{\mu}$ is small close to the ground, 
these altitudes are ignored in the calculation of the effective temperature.
Occasionally, AIRS stops data taking for a few hours for calibration of the instrument. This causes larger gaps in the temperature data. These are filled with temperature data that are interpolated between the previous and following days.

\begin{figure}
    \centering
    \includegraphics[width=0.49\textwidth]{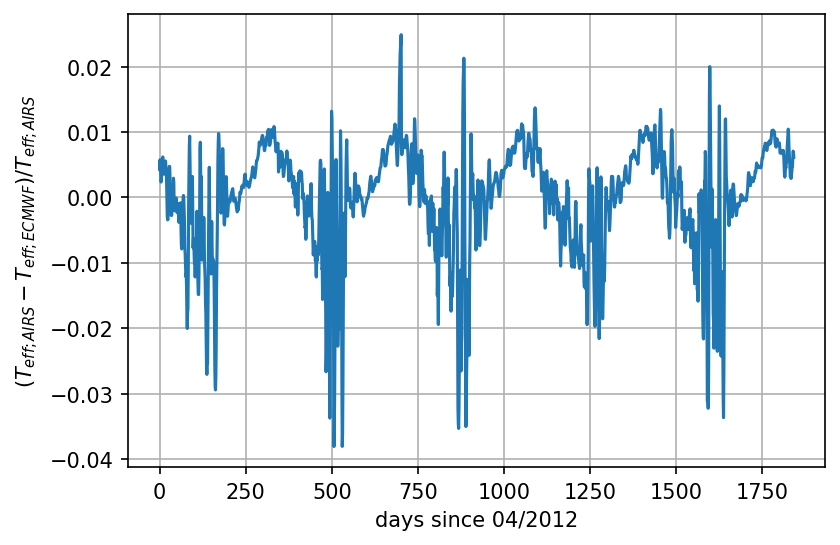}
    \includegraphics[width=0.49\textwidth]{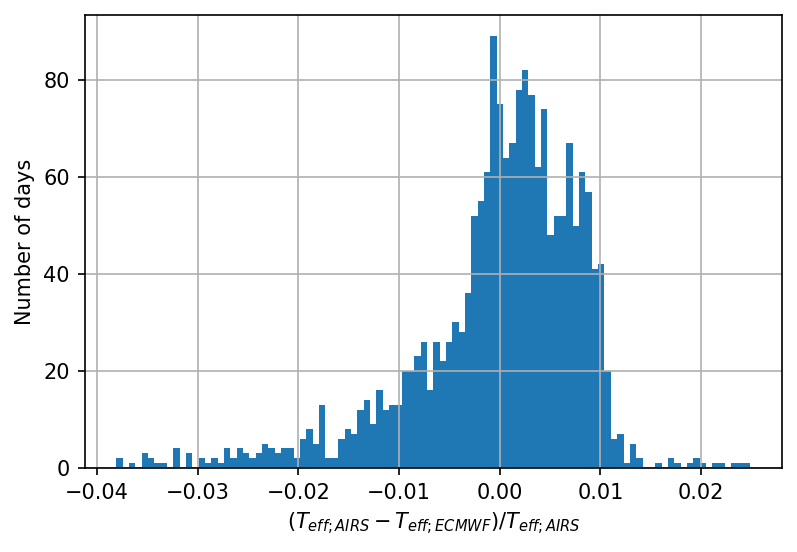}
    \caption{Plots showing the relative difference of the effective temperatures based on the AIRS satellite measurements and the ERA5 data. In the top plot the dependence on time is shown. The largest deviations take place during the spring and fall seasons. The bottom plot shows the distributions of relative differences between the data sets. The mean deviation is \num{5e-5}  and the standard deviation is 0.8\%. }
    \label{fig:ECMWF_comparison}
\end{figure}

For evaluating the accuracy of the AIRS measurement,  a second data set  by the  European Centre for Medium-Range Weather Forecasts (ECMWF) is analyzed~\cite{ERA5data}. The ERA5 data includes AIRS data but also includes measurements from multiple other stations and satellites around the globe, and is thus partly independent. Additionally, the data is combined with atmosphere models and can be evaluated on a fine grid in time (hourly) as well as position on Earth  ($0.25^{\circ} \times 0.25^{\circ}$) and height (37 pressure levels).
Using these temperatures, a secondary set of effective temperatures is calculated and compared to the AIRS result (see Fig.  \ref{fig:ECMWF_comparison}). 
Except for a few larger deviations on the few percent level, differences are generally small
with a mean relative difference that is consistent with zero (\num{5e-5}) and a standard deviation of \SI{0.8}{\percent}.
These studies confirm that the calculation 
of effective temperatures based on the used satellite data is robust and uncertainties are 
smaller than \SI{10}{\percent}  of  the amplitude of the seasonal temperature variation.

\subsection{Atmospheric Depth
Integration \label{sup:integrate}}
\begin{figure*}
    \centering
    \includegraphics[width=0.59\textwidth]{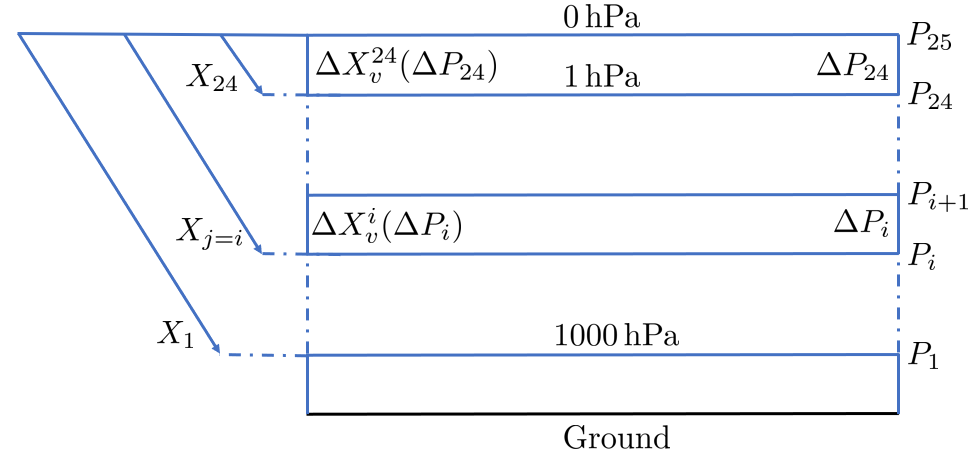}
    \includegraphics[width=0.4\textwidth]{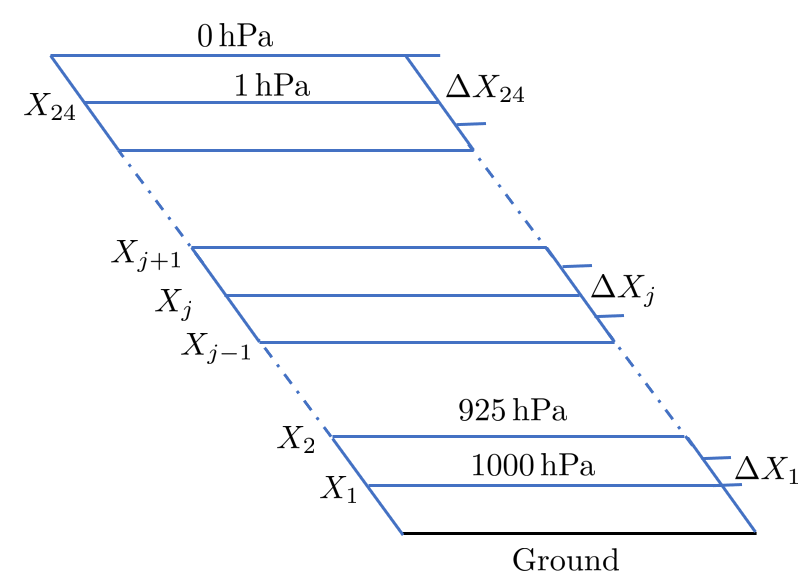}
    \caption{Figure sketching the numerical integration of the pressure levels. The left plot describes the relation between pressure $P_i$, slant depth $X_{\nu}^i$ and atmospheric depth. The right plot describes the definition of $\Delta X_i$.}
    \label{fig:Integration_steps}
\end{figure*}
The  calculation of the effective temperature (Eq.\ref{eq:TeffDef}) proceeds by an integration over the slant atmospheric depth $X$ for the given direction, i.e.\ zenith $\theta $ and azimuth $\varphi$. The depth is discretized for the  altitudes $h_j$ given by the  pressure levels $j$ of the satellite data (see Table \ref{tab:presslevel}). The discretization is  depicted in Fig.  \ref{fig:Integration_steps}.
The slant atmospheric depth for the pressure level $j$ is given by
 \begin{equation}
    X_j(\theta) = \int_{0}^{X_j} dX \approx \sum_{i=j}^{24} \Delta X_i.
\end{equation}
with $\Delta X_i = X_i -X_{i+1} $.
Using the relation between vertical atmospheric depth and slant depth 
$X_{v} = X \frac{dh}{dl} \approx X \cdot \cos \theta^* $  this sum becomes
 \begin{equation}
    X_j(\theta) \approx  \sum_{i=j}^{24} \Delta X_{v,i} \cdot \frac{\Delta l_i}{\Delta h_i} \approx
   \sum_{i=j}^{24}   \frac{\Delta X_{v,i}}{\cos \theta^*}. 
\end{equation}
and with the relation $X_v = \frac{p}{g} $ 
 \begin{equation}
    X_j(\theta) \approx  
   \sum_{i=j}^{24}   \frac{\Delta p_{i}}{ g \cdot \cos \theta^*} 
\end{equation}
with the measured pressure levels $ \Delta p_{i} = p_{i}  - p_{i+1} $.
For the highest pressure level
we assume $P_{25} = \SI{0}{\hecto \pascal}$ which also corresponds to  $X_{25} = \SI{0}{\gram \per \square \centi \metre}$.
For the integration of the effective atmospheric temperature the nominator in Eq. \ref{eq:TeffDef} then becomes
\begin{equation}
    T_{\mathrm{eff}}(\theta 
    ) \approx
   \sum_{i=0}^{24}  \frac{\Delta p_{i}}{ g \cdot \cos \theta^*} \cdot T_i  \cdot \int dE \; P(X_i,E,\theta,T_i) \; A_{\mathrm{eff}}(E,\theta).
  \end{equation}
and the denominator changes correspondingly.
As the temperature values $T_i$ depend on time and direction, also the effective temperature implicitly depends on $\varphi $ and $t$ in addition to the explicit $\theta $ dependence.

Note that the angle $\theta^* $ has to take into account the curvature of Earth, as it defines the local zenith angle of the neutrino production. Its relation to the observed direction $\theta $ can be approximated as 
 \begin{equation}
  \cos \theta^* \approx \sqrt{1 - \left ( \frac{R_E}{R_E+h} \sin \theta \right )^2}
\end{equation}
with the Earth radius $R_E $ and the neutrino production height $h$ in the atmosphere. As $h > 0$, also $\cos \theta^* > 0 $ and diverging terms are avoided in the above calculation.

Integrations over energy and zenith are also approximated by simple sums.
For the energy we use $\int \, dE \approx \sum_i \Delta E_i  $, with typically 50 bins in $\log(E) $ ranging from \SI{100}{\;GeV} to \SI{10}{\;PeV}.
The  angular integration in Eq. \ref{eq:teffint} is approximated as
\begin{equation}
    \int d\Omega = \sum_{i,j}  \Delta \theta_i \sin(\theta_i) \Delta \varphi_j .
\end{equation}
using the  $1^{\circ} \times 1^{\circ}$ grid of the AIRS temperature data.

Note, that for the angular integration of the effective temperature the individual azimuth bins have to be aligned to the time zone of the neutrinos (UTC) as the temperatures are measured in local time. 
To get a singular value of each day, the temperatures are interpolated in each angular bin to a UTC time of 12:00 AM and 12:00 PM by converting the local times to UTC according to the bins in longitude. Details are given in~\cite{MasterThesisSimonHauser}.

\subsection{Systematic Data Splits \label{sup:datasplit}}

\begin{figure}
    \centering
    \includegraphics[width=0.45\textwidth]{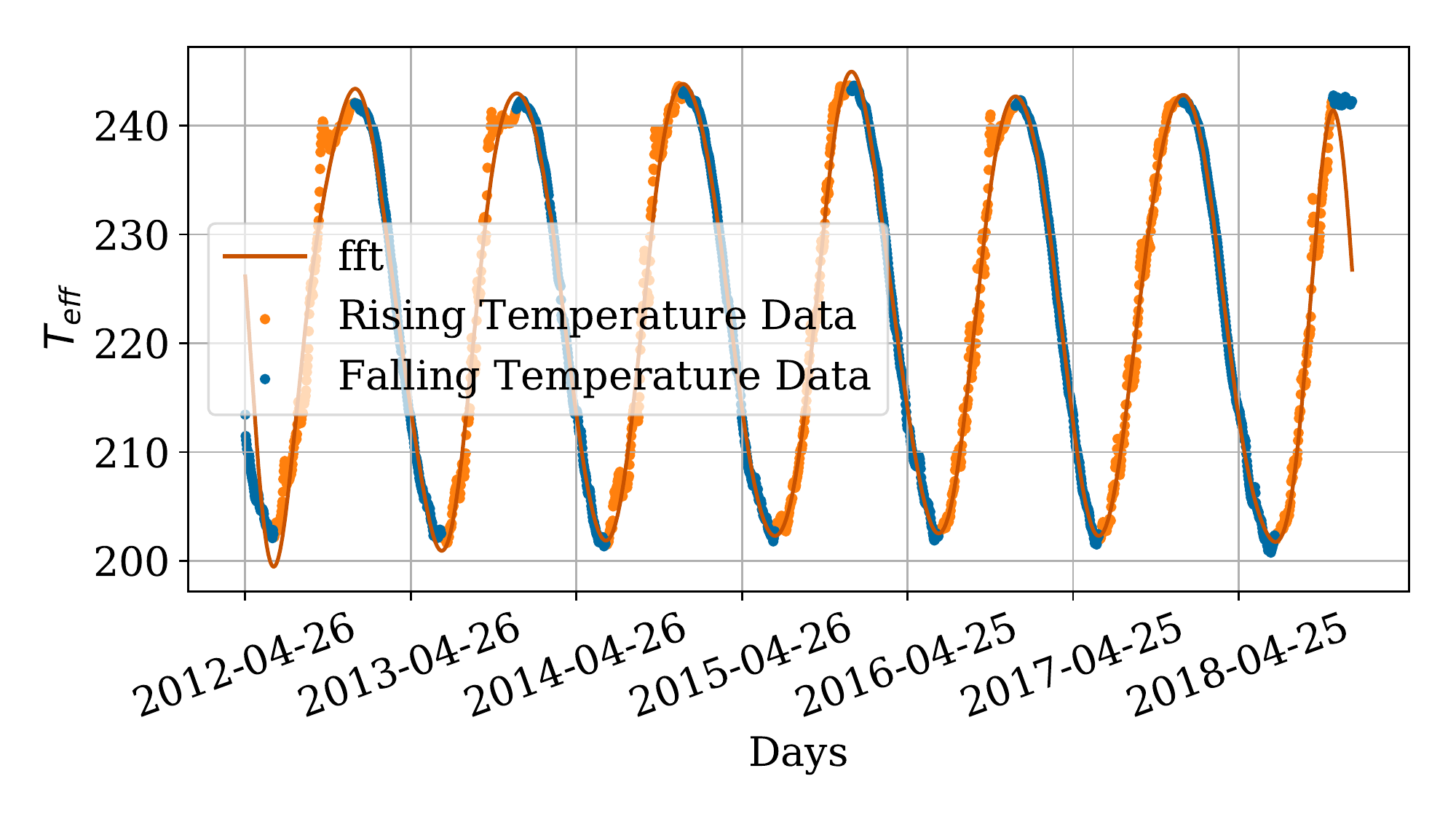}
    \includegraphics[width=0.45\textwidth]{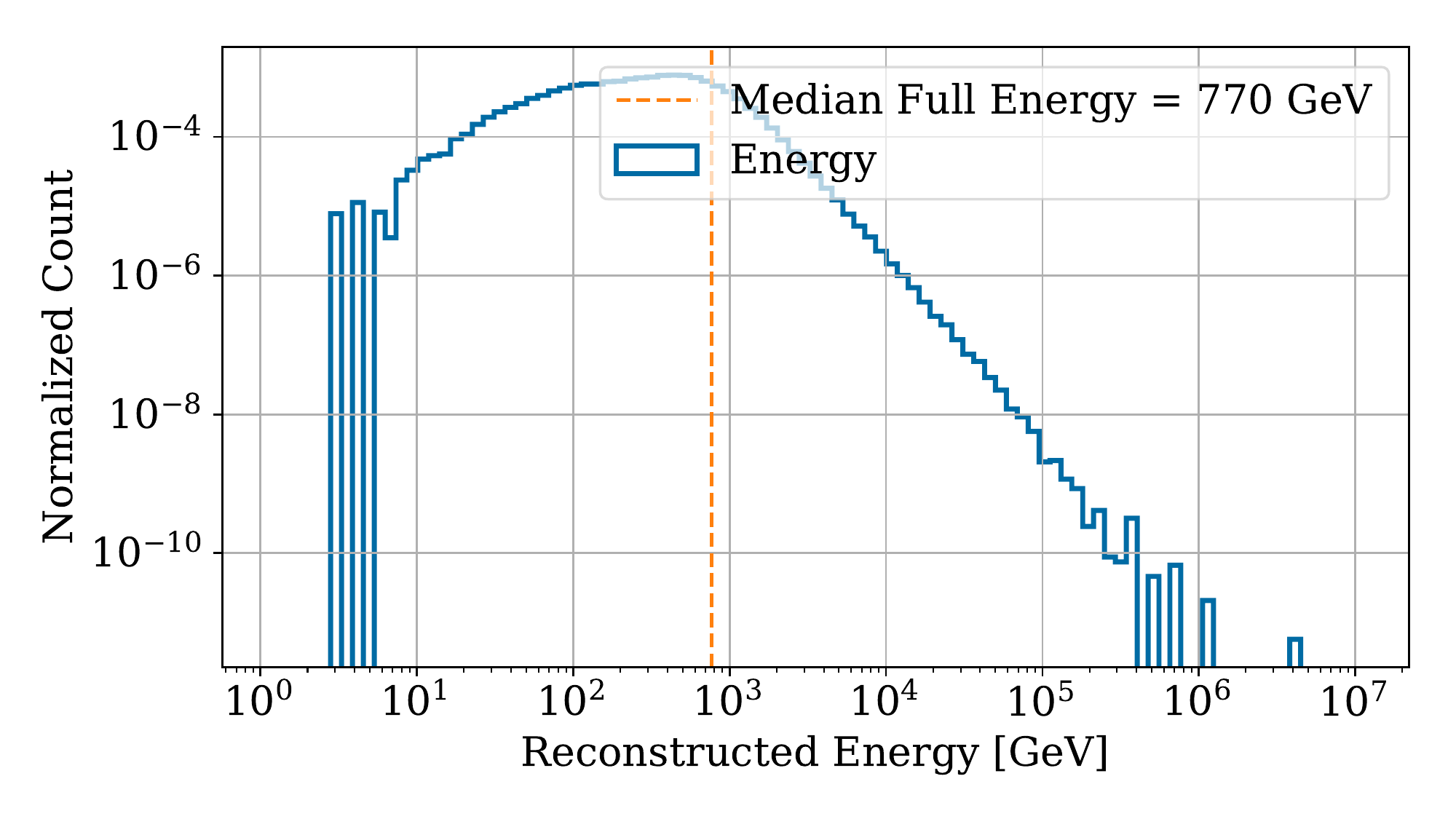}
    \includegraphics[width=0.45\textwidth]{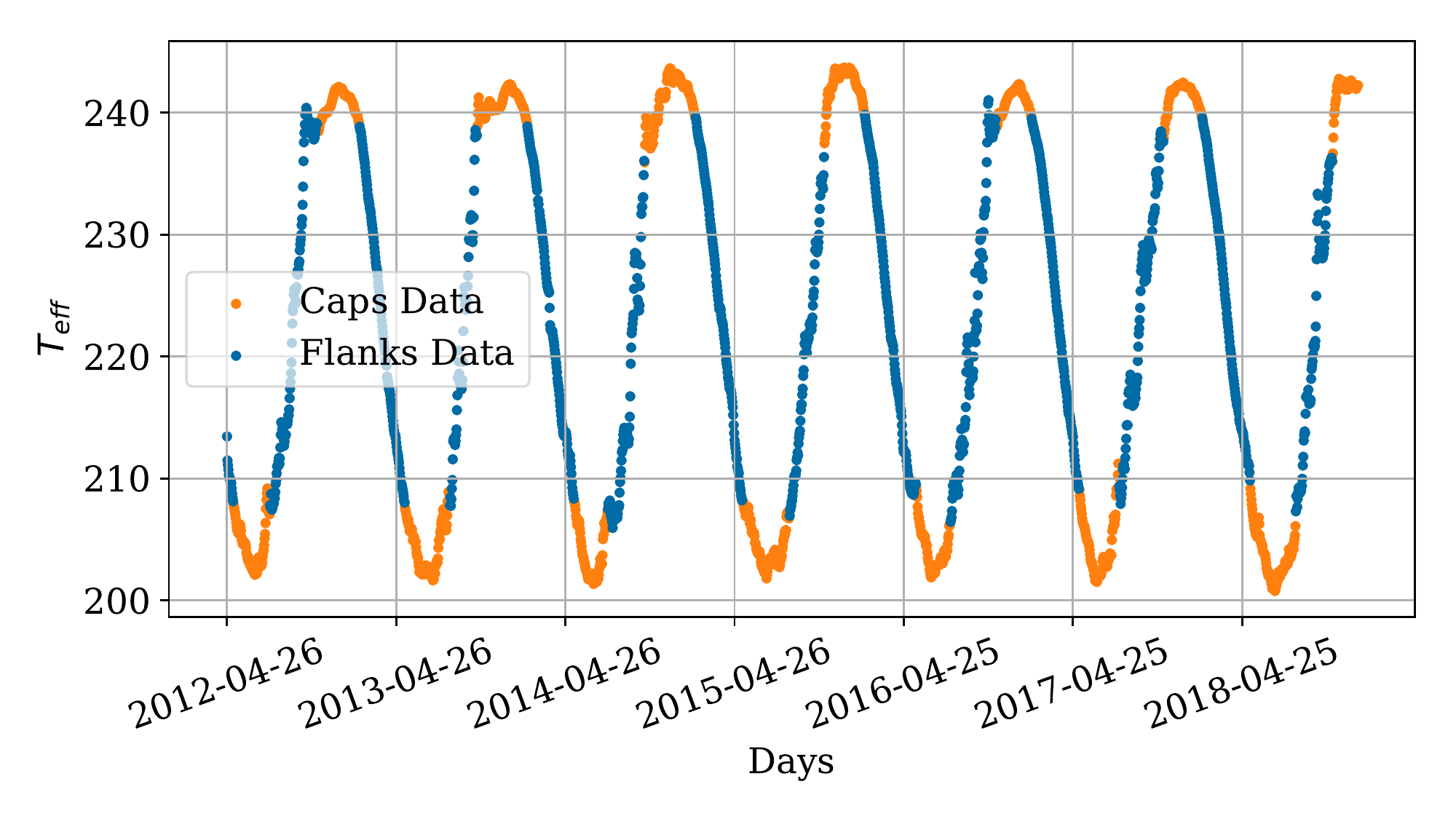}
    \caption{Plots describing the systematic splits applied to the data. On the top, the data is split into day of rising (orange) and falling (blue) temperature. In the middle, the split is done along the median reconstructed energy (truncated energy in \cite{IceCubeEreco}). In the bottom, \SI{50}{\percent} of days of extreme temperatures called caps (orange) are split from the rest which form the flanks data (blue)}
    \label{fig:splits_description}
\end{figure}

The systematic splits of the  data set that are discussed in section \ref{sec:systematics_splits}  are illustrated in Fig.  \ref{fig:splits_description}.

\end{appendix}

\clearpage





\bibliographystyle{spphys}     
\bibliography{references}   

%
%


\end{document}

%% file: authorlist.tex


\onecolumn
\author{R. Abbasi\thanksref{loyola}
\and M. Ackermann\thanksref{zeuthen}
\and J. Adams\thanksref{christchurch}
\and S. K. Agarwalla\thanksref{madisonpac,a}
\and N. Aggarwal\thanksref{edmonton}
\and J. A. Aguilar\thanksref{brusselslibre}
\and M. Ahlers\thanksref{copenhagen}
\and J.M. Alameddine\thanksref{dortmund}
\and N. M. Amin\thanksref{bartol}
\and K. Andeen\thanksref{marquette}
% ...
}
\authorrunning{IceCube Collaboration}
\thankstext{a}{also at Institute of Physics, Sachivalaya Marg, Sainik School Post, Bhubaneswar 751005, India}
\thankstext{b}{also at Department of Space, Earth and Environment, Chalmers University of Technology, 412 96 Gothenburg, Sweden}
\thankstext{c}{also at Earthquake Research Institute, University of Tokyo, Bunkyo, Tokyo 113-0032, Japan}

 \institute{
% ...
Universit{\'e} Libre de Bruxelles, Science Faculty CP230, B-1050 Brussels, Belgium \label{brusselslibre}
% ...
\and Department of Physics, Loyola University Chicago, Chicago, IL 60660, USA \label{loyola}
\and Dept. of Physics and Astronomy, University of Canterbury, Private Bag 4800, Christchurch, New Zealand \label{christchurch}
% ...
\and Niels Bohr Institute, University of Copenhagen, DK-2100 Copenhagen, Denmark \label{copenhagen}
\and Dept. of Physics, TU Dortmund University, D-44221 Dortmund, Germany \label{dortmund}
% ...
\and Dept. of Physics, University of Alberta, Edmonton, Alberta, Canada T6G 2E1 \label{edmonton}
% ...
\and Dept. of Physics and Wisconsin IceCube Particle Astrophysics Center, University of Wisconsin{\textendash}Madison, Madison, WI 53706, USA \label{madisonpac}
% ...
\and Department of Physics, Marquette University, Milwaukee, WI, 53201, USA \label{marquette}
% ...
\and Bartol Research Institute and Dept. of Physics and Astronomy, University of Delaware, Newark, DE 19716, USA \label{bartol}
% ...
\and Deutsches Elektronen-Synchrotron DESY, Platanenallee 6, 15738 Zeuthen, Germany  \label{zeuthen}
}
\maketitle
\twocolumn
